\documentclass[11pt,a4paper]{article}

\usepackage[utf8]{inputenc}
\usepackage[T1]{fontenc}
\usepackage[margin=2.5cm]{geometry}
\usepackage{amsmath,amssymb}
\usepackage{graphicx}
\usepackage{booktabs}
\usepackage{multirow}
\usepackage{array}
\usepackage{caption}
\usepackage{subcaption}
\usepackage{hyperref}
\usepackage{xcolor}
\usepackage{float}
\usepackage{siunitx}
\usepackage{enumitem}
\usepackage{placeins}
\usepackage{pdflscape}

\hypersetup{
  colorlinks=true,
  linkcolor=blue!60!black,
  citecolor=green!40!black,
  urlcolor=blue!60!black,
}

\newcommand{\passk}[1]{\text{Pass@}#1}

\begin{document}

\title{\textbf{Benchmarking API Drift in LLM-Generated Quantum~Code Across Successive SDK Versions}}

\author{%
  Mohammad Arif Rasyidi and Syahirul Faiz \\[2pt]
  \small Department of Computer Science \\
  \small Khalifa University \\
  \small Abu Dhabi, United Arab Emirates \\[2pt]
  \small \texttt{\{100066916, 100065736\}@ku.ac.ae}
}
\date{}
\maketitle

\begin{abstract}
  Large language models can generate plausible quantum code, but it is unclear
  whether they can reliably target the specific software development kit (SDK)
  version requested by the user. We study this problem as \emph{API drift} and
  introduce \textsc{quantum-api-drift}, a benchmark for measuring version
  fidelity, defined here as execution success on the requested SDK version,
  cross-version compatibility, failure modes, and documentation-guided repair in
  LLM-generated quantum SDK code. We instantiate the benchmark with Qiskit, a
  representative quantum SDK that underwent substantial interface changes across
  v0.43, v1.3, and v2.0. We evaluate 17 models on 50 tasks with 3 samples per
  prompt, yielding 450 generated samples and 1{,}350 executions per model.
  Sixteen models are tested in a matched REST API setting with a 1024-token
  output cap, while GPT-5.4 (Codex CLI) is reported separately as a reference
  configuration. Across the 16 matched REST models, diagonal Pass@1 ranges from
  0.02 to 0.85. Claude Opus 4.7 is strongest on v0.43 and v2.0, while Grok 4.20
  is strongest on v1.3 at 0.513. Error profiles differ systematically by model
  strength: weaker models fail mainly with broken imports, while stronger models
  more often reach deprecation-level failures. Documentation-guided repair
  succeeds for 0.19 to 0.59 of repair attempts overall and is consistently much
  more effective for migration to v2.0 than to v1.3. The benchmark artifacts are
  publicly available at \url{https://github.com/arasyi/quantum-api-drift}. These
  results show that version alignment is a distinct evaluation axis for quantum
  code generation and that API drift remains only partly recoverable even with
  migration guidance.
\end{abstract}


\section{Introduction}
\label{sec:intro}

Large language models are increasingly used as programming assistants for
scientific software, including quantum software development kits (SDKs). In
practice, however, users rarely ask for code against an abstract SDK. They ask
for code that must run against a specific installed version. This creates a
failure mode that is easy to recognize but still poorly measured: a model may
generate code that appears sensible, yet targets the wrong API surface for the
environment in which it will be executed. We study this problem as
\emph{API drift} in LLM-generated quantum code.

The problem is broader than any single framework. Quantum software stacks evolve
rapidly, tutorials age quickly, and generated code often sits at the boundary
between static model knowledge and dynamic tool environments. As a result,
version mismatches can arise even when the model appears strong on general code
generation. Despite this practical importance, existing quantum-code evaluation
mostly asks whether a generated solution passes a task, not whether it targets
the requested SDK version or remains robust across version transitions.

This gap sits at the intersection of two mature literatures. In code
generation, execution-based benchmarks commonly evaluate correctness with
Pass@$k$ and related success metrics~\cite{chen2021codex, austin2021program}.
In software evolution, API change is a long-established source of downstream
breakage~\cite{dig2006how, bavota2015how, hora2015api}. Prior work on quantum
code generation, including Qiskit HumanEval, shows that execution-based
evaluation is feasible for quantum programming tasks~\cite{vishwakarma2024qiskithumaneval}.
What remains underexplored is the versioned setting: if a model is explicitly
asked to write code for one SDK release, how often does it actually align with
that release, how often does the code remain executable in other releases, and
how often can version-induced failures be repaired from migration guidance?

To answer these questions, we introduce \textsc{quantum-api-drift}, a benchmark
for measuring version fidelity, defined as execution success on the requested
SDK version, cross-version compatibility, failure modes, and
documentation-guided repair in LLM-generated quantum SDK code. The benchmark is
designed to generalize across quantum SDKs, but in this paper we instantiate it
with Qiskit as a concrete case study. Qiskit is an appropriate test bed because
its transitions from v0.43 to v1.3 and then to v2.0 include visible changes in
imports, execution workflows, primitive usage, backend abstractions, and
deprecation pathways~\cite{qiskit_migration_v1, qiskit_release_1_3,
  qiskit_migration_v2, qiskit_release_2_0}. These changes allow us to study not
only whether models can generate quantum code, but whether they can select the
correct local API idiom for the version named in the prompt.

Our evaluation covers 17 models, 50 tasks, and 3 samples per prompt, yielding
450 generated samples and 1,350 executions per model when each sample is run in
all three environments. Sixteen models are evaluated under a matched REST API
setup with a 1024-token output cap. We also report GPT-5.4 (Codex CLI) as a
reference configuration because it uses a different access pathway and output
budget and therefore should not be interpreted as a like-for-like comparator.

The paper makes three main contributions. First, it introduces a benchmark and
evaluation protocol for studying version alignment in quantum SDK code
generation. Second, it provides a comparative empirical study showing that API
drift is large, structured, and strongly version-dependent, with Qiskit v1.3
emerging as the hardest target in this benchmark. Third, it shows that failure
modes and repairability are not uniform: weaker models fail earlier with broken
imports, stronger models more often produce deprecation-heavy code, and
migration-note prompting helps substantially more for migration to v2.0 than for
migration to v1.3.

Taken together, these findings argue that version alignment should be treated as
a first-class evaluation axis for quantum code generation rather than as a minor
implementation detail. The rest of the paper describes the benchmark design,
reports the empirical results, and discusses the implications for future
quantum-SDK evaluation and tool design.

\section{Methodology}
\label{sec:methodology}

\subsection{Problem Set}

We constructed 50 quantum programming tasks drawn from canonical Qiskit use
cases: circuit construction, gate operations, transpilation, backend
simulation, measurement, and classical-quantum interfacing. The tasks are
sourced from the Qiskit HumanEval benchmark~\cite{vishwakarma2024qiskithumaneval}, from
which we selected a representative 50-problem subset and rewrote both the
prompts and test harnesses to be \emph{SDK-version-agnostic}. Specifically,
the test harnesses avoid API calls that behave differently across versions
(e.g., no \texttt{execute()}, no version-specific primitive constructors),
relying instead on structural assertions such as \texttt{isinstance} checks and
return-value presence tests. This rewriting was necessary to isolate
generation-time API knowledge from test-time execution; however, it introduces
a caveat: for some tasks the semantic correctness of the generated circuit
cannot be guaranteed from the test alone (see Section~\ref{sec:limitations}).
Each task specifies a function signature (\texttt{entry\_point}) and a
lightweight test harness (\texttt{test\_call}).

\subsection{Benchmark Versions}

We avoid transition releases because they mix current behavior with future
deprecation warnings. Instead, we evaluate one stable release from each API
epoch:

\begin{itemize}[leftmargin=*]
  \item \textbf{v0.43} (Qiskit 0.43.3, released Apr 2023): A stable
        mid-series release from the legacy \texttt{qiskit-terra} era. It predates
        the version-number synchronisation and package rename introduced in later
        releases (from v0.45 onward), and largely precedes the major v1.0
        deprecation cycle. It reflects an API surface that is widely represented
        in publicly available Qiskit code examples.

  \item \textbf{v1.3} (Qiskit 1.3.0, released Nov 2024): A later stable
        release within the v1.x series. Under Qiskit's semantic versioning
        guarantees, v1.x releases are intended to avoid breaking changes to the
        public API, so v1.3 and v1.0 expose largely the same interface. We select
        v1.3 as a representative mature state of the v1.x design, where the
        \texttt{SamplerV2}/\texttt{EstimatorV2} primitive workflow is fully
        integrated, the Rust-backed DAGCircuit and Target are established, and
        the deprecation of \texttt{qiskit.pulse} has been formally introduced.

  \item \textbf{v2.0} (Qiskit 2.0.0, released Mar 2025): A major stable
        release introducing the second significant API transition. It removes many
        interfaces deprecated during the v1.x series, including the
        \texttt{execute()} function, BackendV1, \texttt{qiskit.pulse}, and several
        legacy circuit attributes.
\end{itemize}

Together, these versions capture the two major transition points in the Qiskit
API. Each version is evaluated in a separate Conda environment with Python 3.10.

\subsection{Prompt Construction}

Prompts are constructed by prepending a version anchor to each problem
description:
\begin{quote}
  \texttt{Using Qiskit v\{X\}, \{task description\}.}\\
  \texttt{Implement \{entry\_point\}(\{args\}).}
\end{quote}
The anchor explicitly identifies the target SDK version, making the model
responsible for selecting the correct API surface. This anchoring strategy
is consistent with documentation-grounded prompting approaches~\cite{
  peng2023impact, zhou2022docprompting}.

\subsection{Models}

We evaluate seventeen models spanning multiple providers, weight-access policies,
architectural families, and training emphases (Table~\ref{tab:models}). The
open-weight portion of the suite encompasses both dense transformer models
(Gemma 4 31B, GPT-OSS 120B, Qwen3 32B, Qwen3 Coder, Devstral 2, Llama 3.3)
and Mixture-of-Experts (MoE) models (Kimi K2, Nemotron 3 Super,
GLM 4.7 Flash, Llama 4 Scout). The closed-weight portion spans OpenAI
configurations (GPT-5.4 (API), GPT-5.4 Nano, GPT-5.3-codex), Anthropic models
(Claude Opus 4.7, Claude Sonnet 4.6), and Grok 4.20 from xAI. Where primary
technical sources are available, the model-family metadata in
Table~\ref{tab:models} follows official system cards, technical reports, model
cards, or release documentation for Claude 4, GPT-5, GPT-OSS, Grok 4, Kimi K2,
Nemotron 3 Super, Gemma 4, Qwen3, Devstral 2, GLM-4.7, Llama 3.3, and Llama
4~\cite{anthropic_claude4_system_card, openai_gpt5_system_card,
  openai_gpt_oss, xai_grok4, kimi_k2_techreport, nemotron3super_techreport,
  gemma4_modelcard, qwen3_techreport, devstral2_modelcard, glm47_docs,
  llama33_modelcard, llama4_modelcard}.

\textbf{GPT-5.4 (Codex CLI) as a reference only.} GPT-5.4 (Codex CLI) is not directly
comparable to the other 16 models because it is accessed through the Codex CLI
and is not subject to the same 1024-token output cap. We therefore report it as
a reference rather than include it in matched REST-API comparisons. We use the
labels \emph{GPT-5.4 (API)} and \emph{GPT-5.4 (Codex CLI)} throughout to
distinguish the matched REST-API configuration from the Codex CLI configuration
without implying that they are different public model families.

All models were accessed with temperature $T = 0.8$ and a maximum of
1024 output tokens. We generate $K = 3$ independent samples per prompt,
yielding 150 executions per (model, prompted-version) cell in the drift
matrix.

Training cutoff is another relevant source of variation. Qiskit v1.3 and v2.0
mark distinct API epochs documented in the corresponding migration and release
materials~\cite{qiskit_migration_v1, qiskit_release_1_3, qiskit_migration_v2, qiskit_release_2_0}. Models trained before those
releases may have limited direct exposure to the relevant APIs. We examine this in
Section~\ref{sec:cutoff}.

\begin{table}[htbp]
  \centering
  \caption{Models evaluated in this study. ``Arch.'' is Dense or MoE; for MoE
    models the parameter column shows total\,/\,active at inference.
    ``K.C.'' = knowledge cutoff; ``?'' = undisclosed.
    $^\dagger$Hybrid LatentMoE with interleaved Mamba-2 and attention layers.
    $^\ddagger$Reference model only: same base model family as GPT-5.4 (API), but
    accessed through Codex CLI with reasoning augmentation and no output token
    limit, so it is not directly comparable to the sixteen chat-completion models.
    ``Vendor'' denotes the organization that developed the model; ``Access''
    identifies the platform used in our experiments. Model metadata are taken from
    official primary sources where available~\cite{anthropic_claude4_system_card,
      openai_gpt5_system_card, openai_gpt_oss, xai_grok4, kimi_k2_techreport,
      nemotron3super_techreport, gemma4_modelcard, qwen3_techreport,
      devstral2_modelcard, glm47_docs, llama33_modelcard, llama4_modelcard}.}
  \label{tab:models}
  \resizebox{\textwidth}{!}{%
    \setlength{\tabcolsep}{4pt}
    \begin{tabular}{llclllll}
      \toprule
      \textbf{Label}                 & \textbf{Model ID}      & \textbf{Wts} & \textbf{Arch.} & \textbf{Params (tot.\,/\,act.)} & \textbf{K.C.} & \textbf{Vendor} & \textbf{Access} \\
      \midrule
      \multicolumn{8}{l}{\textit{Codex CLI reference (unlimited output tokens)}}                                                                                                    \\
      GPT-5.4 (Codex CLI)$^\ddagger$ & gpt-5.4-medium         & Closed       & -              & undisclosed                     & Aug 2025      & OpenAI          & Codex CLI       \\
      \midrule
      \multicolumn{8}{l}{\textit{REST API models, 1024-token output cap}}                                                                                                           \\
      Claude Opus 4.7                & claude-opus-4-7        & Closed       & -              & undisclosed                     & Apr 2025      & Anthropic       & API             \\
      GPT-5.3-codex                  & gpt-5.3-codex          & Closed       & -              & undisclosed                     & Aug 2025      & OpenAI          & API             \\
      GPT-5.4 (API)                  & gpt-5.4                & Closed       & -              & undisclosed                     & Aug 2025      & OpenAI          & API             \\
      Grok 4.20                      & grok-4-20              & Closed       & -              & undisclosed                     & Apr 2025      & xAI             & API             \\
      Claude Sonnet 4.6              & claude-sonnet-4-6      & Closed       & -              & undisclosed                     & Apr 2025      & Anthropic       & API             \\
      GPT-5.4 Nano                   & gpt-5.4-nano           & Closed       & -              & undisclosed                     & Aug 2025      & OpenAI          & API             \\
      Kimi K2                        & kimi-k2-instruct       & Open         & MoE            & 1\,T / 32B                      & Apr 2025      & Moonshot AI     & API             \\
      Nemotron 3 Super               & nemotron-3-super-120b  & Open         & MoE$^\dagger$  & 120B / 12B                      & Mar 2025      & NVIDIA          & API             \\
      GPT-OSS 120B                   & gpt-oss-120b           & Open         & Dense          & 120B                            & ?             & OpenAI          & API             \\
      Gemma 4 31B                    & gemma-4-31b-it         & Open         & Dense          & 31B                             & Jan 2025      & Google          & API             \\
      Qwen3 Coder                    & qwen3-coder            & Open         & Dense          & undisclosed                     & ?             & Alibaba         & API             \\
      Devstral 2                     & devstral-2-123b        & Open         & Dense          & 123B                            & ?             & Mistral AI      & API             \\
      Qwen3 32B                      & qwen3-32b              & Open         & Dense          & 32B                             & Oct 2024      & Alibaba         & API             \\
      GLM 4.7 Flash                  & glm-4.7-flash          & Open         & MoE            & 30B / 3B                        & ?             & Z.ai            & API             \\
      Llama 3.3                      & llama-3.3-70b-instruct & Open         & Dense          & 70B                             & Dec 2023      & Meta            & API             \\
      Llama 4 Scout                  & llama-4-scout-17b-16e  & Open         & MoE            & 109B / 17B                      & Aug 2024      & Meta            & API             \\
      \bottomrule
    \end{tabular}%
  }
\end{table}

\subsection{Execution Protocol}

Each generated code snippet is executed in \emph{all three} Qiskit
environments, regardless of which version it was prompted for. A test
runner wraps the snippet in a try/except block and classifies execution
outcomes into three categories:
\begin{itemize}[leftmargin=*]
  \item \textbf{Pass}: Code executes without exception and the test assertion
        succeeds.
  \item \textbf{Soft-fail}: Code triggers a \texttt{DeprecationWarning}
        but otherwise executes correctly. This represents warning-level drift.
  \item \textbf{Hard-fail}: Code raises an exception (ImportError, AttributeError,
        TypeError, etc.). This represents breaking drift.
\end{itemize}
Execution is sandboxed via \texttt{conda run} with a 30-second timeout.

\subsection{Drift Matrix}

For each model, we compute a $3 \times 3$ drift matrix $D$, where entry
$D[s][t]$ is the fraction of samples prompted for version $s$ that pass
execution in version $t$. The diagonal $D[v][v]$ measures \emph{version
  fidelity}, that is, target-version execution success for code prompted on
version $v$. Off-diagonal entries measure \emph{cross-version compatibility}.

\subsection{Pass@k Evaluation}

With $n = 3$ samples per prompt, $\passk{1}$ for prompt $i$ is the fraction of
individual samples that pass, while $\passk{3}$ indicates whether at least one
of the three samples for prompt $i$ passes. Following~\cite{chen2021codex}, we
compute the per-prompt unbiased estimator
\begin{equation}
  \passk{k}_i = 1 - \frac{\binom{n-c_i}{k}}{\binom{n}{k}}
\end{equation}
where $c_i$ is the number of passing samples for prompt $i$. We then report the
benchmark-level score as the macro-average across all $N$ prompts:
\begin{equation}
  \passk{k}_{\mathrm{benchmark}} = \frac{1}{N}\sum_{i=1}^{N} \passk{k}_i
\end{equation}
All proportions are reported with 95\% Wilson score confidence
intervals~\cite{wilson1927probable}, which are well-behaved at extreme
proportions.

\subsection{Error Taxonomy}

We classify each failure by its top-level Python exception type. The primary
categories observed are: \texttt{ImportError} / \texttt{ModuleNotFoundError},
\texttt{DeprecationWarning}, \texttt{AttributeError}, \texttt{TypeError},
\texttt{AssertionError}, \texttt{QiskitError}, \texttt{CircuitError}, and
\texttt{ValueError}.

\subsection{Documentation-Augmented Repair}

For each snippet that (i) passes in its prompted version and (ii) fails in at
least one newer version, we attempt automated repair. The repair prompt
provides the model with:
\begin{enumerate}[leftmargin=*]
  \item The failing code and its error trace.
  \item The relevant Qiskit migration notes (plain-text summaries of breaking
        changes for the relevant version transition).
  \item The original task description.
\end{enumerate}
The repaired snippet is executed in the target environment to verify success.
Repair success rate is computed as the fraction of attempted repairs that
yield a passing execution.

\section{Results}
\label{sec:results}

\subsection{Version Fidelity}

Table~\ref{tab:fidelity} reports version fidelity, that is, the diagonal of the
drift matrix. Among the 16 matched REST models, Claude~Opus~4.7 is best on
v0.43 ($\passk{1}=0.72$) and v2.0 ($0.853$), while Grok~4.20 is best on v1.3
($0.513$). The main empirical pattern is the weak v1.3 column: every REST model
falls between $0.02$ and $0.513$ on that target.

\begin{table}[htbp]
  \centering
  \caption{Version fidelity: Pass@1 and Pass@3 for each model on its target
    version (diagonal of drift matrix $D[v][v]$). Wilson 95\% CIs shown in
    brackets. Best per-column values in \textbf{bold} (across all models).
    Models are grouped as: Codex CLI reference, closed REST-API, open-weight REST-API.}
  \label{tab:fidelity}
  \setlength{\tabcolsep}{2.4pt}
  \scriptsize
  \begin{tabular}{l|cc|cc|cc}
    \toprule
                                   & \multicolumn{2}{c|}{\textbf{v0.43}} & \multicolumn{2}{c|}{\textbf{v1.3}} & \multicolumn{2}{c}{\textbf{v2.0}}                                                                                     \\
    \textbf{Model}                 & \passk{1}                           & \passk{3}                          & \passk{1}                         & \passk{3}                 & \passk{1}                 & \passk{3}                 \\
    \midrule
    \multicolumn{7}{l}{\textit{Codex CLI reference}}                                                                                                                                                                                  \\
    GPT-5.4 (Codex CLI)$^\ddagger$ & .673 [.595,.743]                    & .760 [.626,.857]                   & .253 [.190,.328]                  & .340 [.224,.478]          & .820 [.751,.873]          & \textbf{.900 [.786,.957]} \\
    \midrule
    \multicolumn{7}{l}{\textit{Closed REST-API models}}                                                                                                                                                                               \\
    Claude Opus 4.7                & \textbf{.720 [.643,.786]}           & .760 [.626,.857]                   & .433 [.357,.513]                  & .460 [.330,.596]          & \textbf{.853 [.788,.901]} & .880 [.762,.944]          \\
    GPT-5.3-codex                  & .647 [.564,.722]                    & .708 [.568,.818]                   & .358 [.282,.441]                  & .426 [.295,.567]          & .798 [.719,.860]          & .818 [.680,.905]          \\
    GPT-5.4 (API)                  & .693 [.615,.762]                    & \textbf{.800 [.670,.888]}          & .267 [.202,.343]                  & .320 [.208,.458]          & .700 [.622,.768]          & .840 [.715,.917]          \\
    Grok 4.20                      & .587 [.507,.662]                    & .640 [.501,.759]                   & \textbf{.513 [.434,.592]}         & \textbf{.560 [.423,.688]} & .633 [.554,.706]          & .680 [.542,.792]          \\
    Claude Sonnet 4.6              & .607 [.527,.681]                    & .680 [.542,.792]                   & .213 [.155,.286]                  & .240 [.143,.374]          & .700 [.622,.768]          & .780 [.648,.872]          \\
    GPT-5.4 Nano                   & .360 [.288,.439]                    & .520 [.385,.652]                   & .227 [.167,.300]                  & .360 [.241,.499]          & .400 [.325,.480]          & .560 [.423,.688]          \\
    \midrule
    \multicolumn{7}{l}{\textit{Open-weight REST-API models}}                                                                                                                                                                          \\
    Kimi K2                        & .567 [.487,.643]                    & .760 [.626,.857]                   & .420 [.344,.500]                  & .540 [.404,.670]          & .500 [.421,.579]          & .660 [.522,.776]          \\
    Nemotron 3 Super               & .487 [.408,.566]                    & .640 [.501,.759]                   & .356 [.283,.435]                  & .520 [.385,.652]          & .493 [.414,.573]          & .660 [.522,.776]          \\
    GPT-OSS 120B                   & .517 [.437,.596]                    & .660 [.522,.776]                   & .347 [.275,.426]                  & .460 [.330,.596]          & .427 [.350,.507]          & .580 [.442,.706]          \\
    Gemma 4 31B                    & .427 [.350,.507]                    & .520 [.385,.652]                   & .133 [.088,.197]                  & .160 [.083,.285]          & .473 [.395,.553]          & .620 [.482,.741]          \\
    Qwen3 Coder                    & .413 [.338,.493]                    & .580 [.442,.706]                   & .287 [.220,.364]                  & .380 [.259,.518]          & .313 [.245,.391]          & .460 [.330,.596]          \\
    Devstral 2                     & .369 [.296,.449]                    & .580 [.442,.706]                   & .288 [.220,.366]                  & .420 [.294,.558]          & .331 [.260,.410]          & .480 [.348,.615]          \\
    Qwen3 32B                      & .373 [.300,.453]                    & .500 [.366,.634]                   & .187 [.132,.257]                  & .300 [.191,.438]          & .293 [.226,.371]          & .420 [.294,.558]          \\
    GLM 4.7 Flash                  & .233 [.173,.307]                    & .520 [.385,.652]                   & .207 [.150,.278]                  & .360 [.241,.499]          & .247 [.185,.321]          & .400 [.276,.538]          \\
    Llama 3.3                      & .353 [.281,.433]                    & .460 [.330,.596]                   & .040 [.018,.085]                  & .060 [.021,.162]          & .127 [.083,.189]          & .200 [.112,.330]          \\
    Llama 4 Scout                  & .133 [.088,.197]                    & .220 [.128,.352]                   & .020 [.007,.057]                  & .040 [.011,.135]          & .067 [.037,.118]          & .160 [.083,.285]          \\
    \bottomrule
  \end{tabular}
\end{table}

\begin{figure}[htbp]
  \centering
  \includegraphics[width=\textwidth]{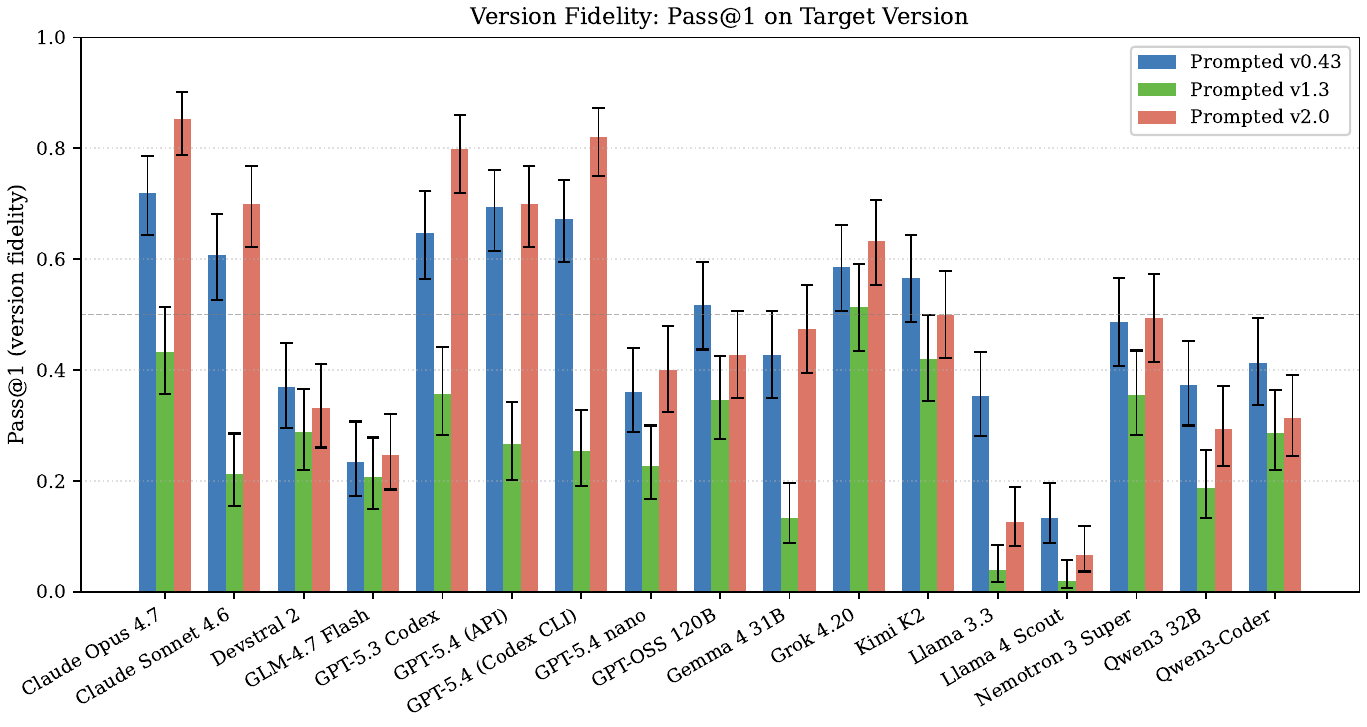}
  \caption{Version fidelity (Pass@1 on the target version, i.e., diagonal of the
    drift matrix) for all models and all three Qiskit versions. Error bars show
    95\% Wilson confidence intervals. The v1.3 column is consistently the weakest
    across models.}
  \label{fig:version_fidelity}
\end{figure}

Figure~\ref{fig:version_fidelity} shows the same result visually. v1.3 is the
hardest target across the full suite, Grok~4.20 is the only model above 0.50 on
that version, and Llama~4~Scout is lowest across all three diagonals.

\subsection{Drift Heatmaps}

\begin{figure}[htbp]
  \centering
  \includegraphics[width=\textwidth]{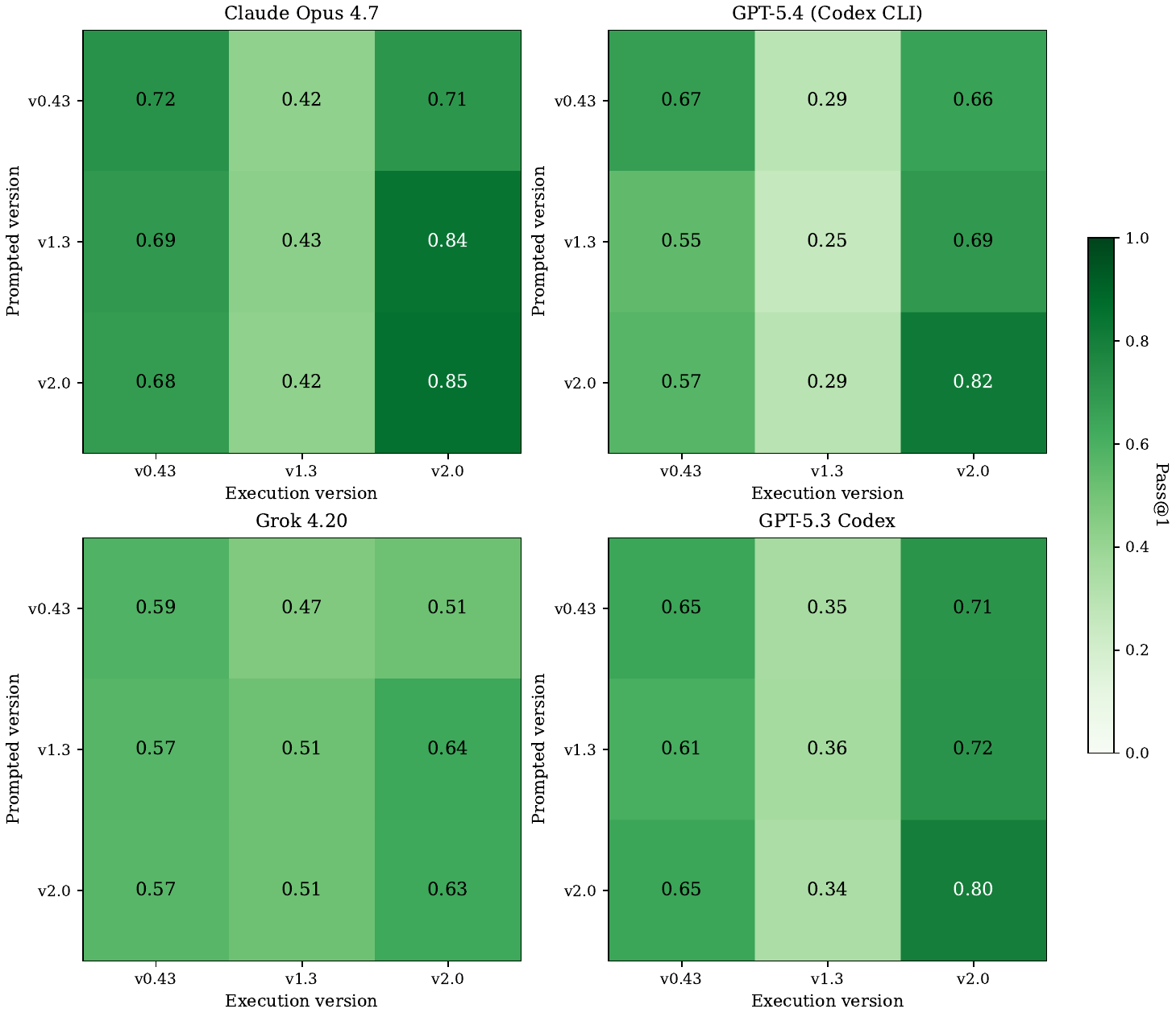}
  \caption{Drift matrices (Pass@1) for the four strongest models. Rows denote
    the prompted Qiskit version; columns denote the execution environment. The
    diagonal represents version fidelity. Low values in the middle column (v1.3)
    are a consistent finding across models.}
  \label{fig:drift_heatmaps}
\end{figure}

Figure~\ref{fig:drift_heatmaps} moves from diagonal fidelity to the full cross-
version picture by showing drift matrices for four representative models. The
shared pattern is clear: performance is strongest on v0.43 and v2.0, but drops
in the v1.3 execution column. Grok~4.20 is the most stable of the matched REST
models because its matrix varies less across versions, whereas Claude~Opus~4.7
and GPT-5.3-codex are stronger overall but still show a marked v1.3 dip.
GPT-5.4 (Codex CLI) follows the same pattern and is included only as a reference
because of its different interface. Among the highlighted models, GPT-5.4
(Codex CLI) shows an especially uneven profile, with high fidelity on v0.43 and
v2.0 but substantially lower fidelity on v1.3.

\subsection{Forward Compatibility Drop}

\begin{figure}[htbp]
  \centering
  \includegraphics[width=\textwidth]{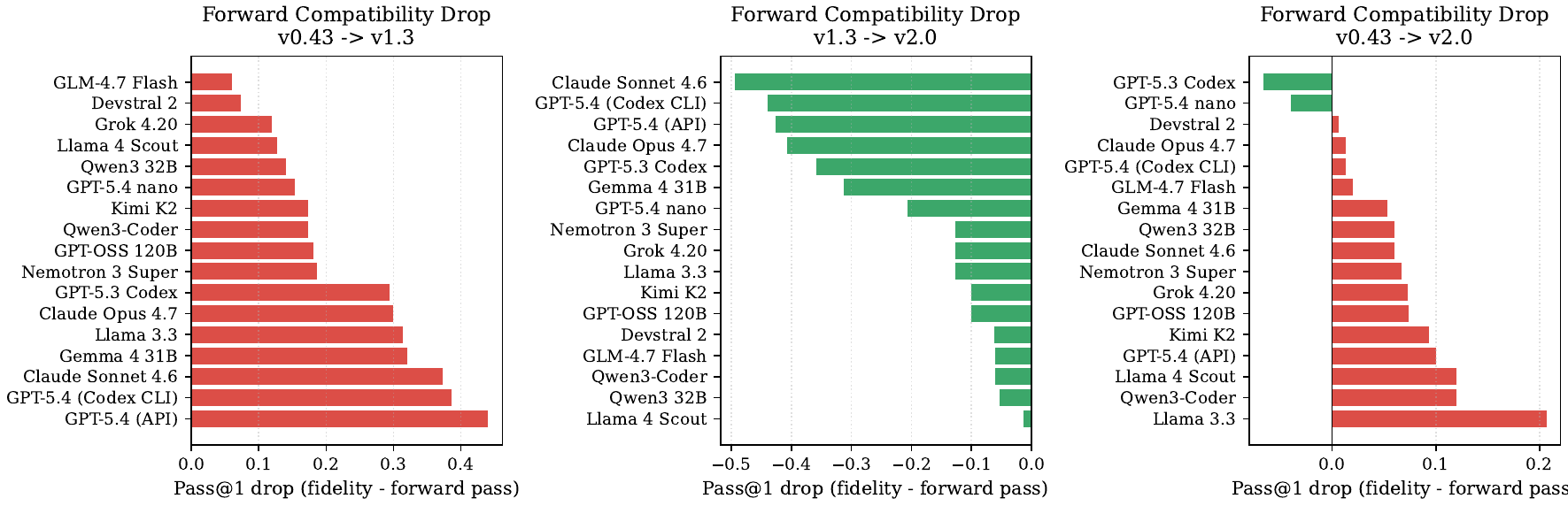}
  \caption{Forward compatibility drop: reduction in Pass@1 when code is executed
    in a newer version than its target. Positive values indicate degradation;
    negative values indicate the code runs \emph{better} in the newer version
    (a sign the model ignores version anchors and generates modern code regardless).}
  \label{fig:forward_drop}
\end{figure}

Figure~\ref{fig:forward_drop} isolates a different aspect of the same story: the
\emph{forward compatibility drop}, defined as the reduction in Pass@1 when code
generated for version $s$ is executed in the newest version $t > s$. The left
panel shows the v0.43 $\to$ v2.0 drop, and the right panel shows the v1.3
$\to$ v2.0 drop.

A few models show only a small forward drop from v0.43 to v2.0, which suggests
that they sometimes generate newer idioms even when prompted for an older
version. In contrast, the v1.3 to v2.0 transition is much less robust across the
suite, reinforcing the view that v1.3 behaves as a difficult middle target.

GPT-5.4 (Codex CLI) illustrates this asymmetry especially clearly. Its
diagonal \passk{1} is strong on v0.43 (0.673) and v2.0 (0.820), but much
lower on v1.3 (0.253). At the same time, when prompted for v1.3 and
executed on v2.0, its \passk{1} rises to 0.693, which substantially
exceeds its v1.3 diagonal score. This pattern suggests that the system
often converges to newer Qiskit idioms even when the prompt specifies the
intermediate version. We therefore interpret GPT-5.4 (Codex CLI) as an
example of strong coding capability combined with uneven version fidelity,
rather than as uniformly robust performance across all requested targets.

\subsection{Error Taxonomy}

\begin{figure}[htbp]
  \centering
  \includegraphics[width=\textwidth]{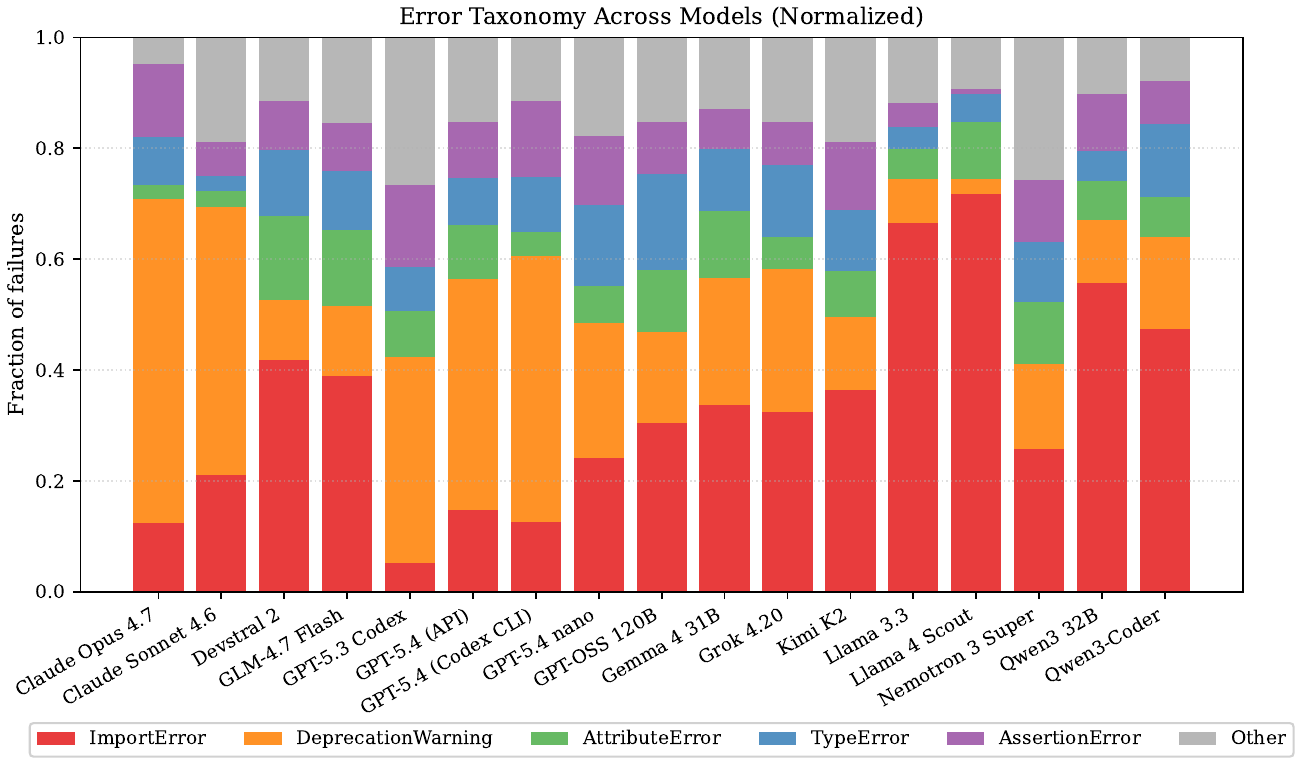}
  \caption{Normalized error taxonomy across models. Weaker models are
    dominated by \texttt{ImportError}, indicating structural module confusion.
    Stronger models shift toward \texttt{DeprecationWarning}, suggesting
    version-lagged but structurally correct code. Error types shown are those
    with the highest aggregate frequency.}
  \label{fig:error_taxonomy}
\end{figure}

Figure~\ref{fig:error_taxonomy} shows two broad failure regimes. Lower-fidelity
models are dominated by \texttt{ImportError}, which points to broken module or
symbol references. Higher-fidelity models shift toward
\texttt{DeprecationWarning}, which suggests that the code structure is often
correct but tied to older interfaces. Grok~4.20 sits between these two regimes.

Table~\ref{tab:error_counts} reports raw error counts for the five most
frequent error types per model, providing absolute scale alongside the
normalized view.

\begin{table}[htbp]
  \centering
  \caption{Raw counts of the 5 most frequent error types per model across
    all executions. Total failures include all error types, not only those shown.
    Models grouped by access type then sorted by total failures (ascending).}
  \label{tab:error_counts}
  \setlength{\tabcolsep}{4pt}
  \begin{tabular}{lrrrrrr}
    \toprule
    \textbf{Model}                 & \textbf{ImportErr} & \textbf{DeprecWarn} & \textbf{TypeError} & \textbf{AssertErr} & \textbf{AttrErr} & \textbf{Total} \\
    \midrule
    \multicolumn{7}{l}{\textit{Codex CLI reference}}                                                                                                        \\
    GPT-5.4 (Codex CLI)$^\ddagger$ & 79                 & 303                 & 62                 & 87                 & 27               & 630            \\
    \midrule
    \multicolumn{7}{l}{\textit{Closed REST-API}}                                                                                                            \\
    Claude Opus 4.7                & 60                 & 284                 & 42                 & 64                 & 12               & 485            \\
    GPT-5.3-codex                  & 26                 & 189                 & 40                 & 75                 & 42               & 507            \\
    Grok 4.20                      & 194                & 155                 & 78                 & 47                 & 34               & 599            \\
    GPT-5.4 (API)                  & 96                 & 270                 & 55                 & 66                 & 63               & 649            \\
    Claude Sonnet 4.6              & 147                & 335                 & 18                 & 43                 & 21               & 695            \\
    GPT-5.4 Nano                   & 218                & 220                 & 133                & 111                & 59               & 902            \\
    \midrule
    \multicolumn{7}{l}{\textit{Open-weight REST-API}}                                                                                                       \\
    Kimi K2                        & 254                & 91                  & 77                 & 85                 & 58               & 697            \\
    Nemotron 3 Super               & 197                & 118                 & 83                 & 86                 & 85               & 766            \\
    GPT-OSS 120B                   & 233                & 127                 & 133                & 72                 & 85               & 767            \\
    Devstral 2                     & 363                & 94                  & 104                & 76                 & 132              & 869            \\
    Qwen3 Coder                    & 426                & 149                 & 117                & 70                 & 66               & 899            \\
    Gemma 4 31B                    & 306                & 209                 & 102                & 66                 & 110              & 910            \\
    Qwen3 32B                      & 539                & 109                 & 53                 & 100                & 68               & 967            \\
    GLM 4.7 Flash                  & 403                & 130                 & 110                & 90                 & 143              & 1035           \\
    Llama 3.3                      & 738                & 88                  & 45                 & 47                 & 60               & 1110           \\
    Llama 4 Scout                  & 901                & 35                  & 63                 & 12                 & 129              & 1256           \\
    \bottomrule
  \end{tabular}
\end{table}

\subsection{Repair Success Rates}

Table~\ref{tab:repair} shows that repair success differs widely by model, from
0.19 to 0.59 across the 16 matched REST models. The most consistent pattern is
not the ranking but the target asymmetry: repair toward v2.0 is much easier than
repair toward v1.3. Several strong models reach 1.00 for repairs to v2.0, while
multiple models are near zero for repairs to v1.3. This conclusion is directly
supported by the table and does not depend on any single model.

\begin{table}[htbp]
  \centering
  \caption{Repair success rates (fraction of drift failures successfully repaired).
    95\% Wilson CIs in brackets. ``$n$'' is the number of repair attempts.
    Models grouped by access type then sorted by overall repair rate (descending).}
  \label{tab:repair}
  \setlength{\tabcolsep}{3.5pt}
  \begin{tabular}{lrr@{\,}lr@{\,}lr@{\,}l}
    \toprule
    \textbf{Model}      & $n$ & \multicolumn{2}{c}{\textbf{Overall}} & \multicolumn{2}{c}{\textbf{$\to$ v1.3}} & \multicolumn{2}{c}{\textbf{$\to$ v2.0}}                                       \\
    \midrule
    \multicolumn{8}{l}{\textit{Codex CLI reference}}                                                                                                                                           \\
    GPT-5.4 (Codex CLI) & 85  & .341                                 & [.249, .447]                            & .152                                    & [.084, .257] & 1.000 & [.832, 1.00] \\
    \midrule
    \multicolumn{8}{l}{\textit{Closed REST-API}}                                                                                                                                               \\
    Grok 4.20           & 44  & .591                                 & [.444, .723]                            & .292                                    & [.149, .492] & .950  & [.764, .991] \\
    Kimi K2             & 51  & .490                                 & [.359, .623]                            & .333                                    & [.192, .512] & .714  & [.500, .862] \\
    Qwen3 Coder         & 52  & .442                                 & [.316, .577]                            & .241                                    & [.122, .421] & .696  & [.491, .844] \\
    GPT-5.4 Nano        & 38  & .342                                 & [.212, .501]                            & .241                                    & [.122, .421] & .667  & [.354, .879] \\
    GPT-5.3-codex       & 54  & .315                                 & [.207, .447]                            & .229                                    & [.133, .365] & 1.000 & [.610, 1.00] \\
    Claude Opus 4.7     & 65  & .308                                 & [.209, .428]                            & .118                                    & [.055, .234] & 1.000 & [.785, 1.00] \\
    GPT-5.4 (API)       & 97  & .278                                 & [.199, .375]                            & .014                                    & [.002, .076] & 1.000 & [.871, 1.00] \\
    Claude Sonnet 4.6   & 83  & .277                                 & [.192, .382]                            & .049                                    & [.017, .135] & .909  & [.722, .975] \\
    \midrule
    \multicolumn{8}{l}{\textit{Open-weight REST-API}}                                                                                                                                          \\
    GPT-OSS 120B        & 42  & .500                                 & [.355, .645]                            & .310                                    & [.173, .492] & .923  & [.667, .986] \\
    Devstral 2          & 34  & .412                                 & [.264, .578]                            & .150                                    & [.052, .360] & .786  & [.524, .924] \\
    Qwen3 32B           & 36  & .389                                 & [.248, .551]                            & .208                                    & [.092, .405] & .750  & [.468, .911] \\
    Nemotron 3 Super    & 44  & .386                                 & [.257, .534]                            & .241                                    & [.122, .421] & .667  & [.417, .848] \\
    GLM 4.7 Flash       & 21  & .333                                 & [.172, .546]                            & .231                                    & [.082, .503] & .500  & [.215, .785] \\
    Gemma 4 31B         & 74  & .311                                 & [.217, .423]                            & .019                                    & [.003, .101] & 1.000 & [.851, 1.00] \\
    Llama 4 Scout       & 37  & .243                                 & [.134, .401]                            & .000                                    & [.000, .168] & .500  & [.290, .710] \\
    Llama 3.3           & 78  & .192                                 & [.120, .293]                            & .000                                    & [.000, .076] & .484  & [.320, .652] \\
    \bottomrule
  \end{tabular}
\end{table}

\begin{figure}[htbp]
  \centering
  \includegraphics[width=\textwidth]{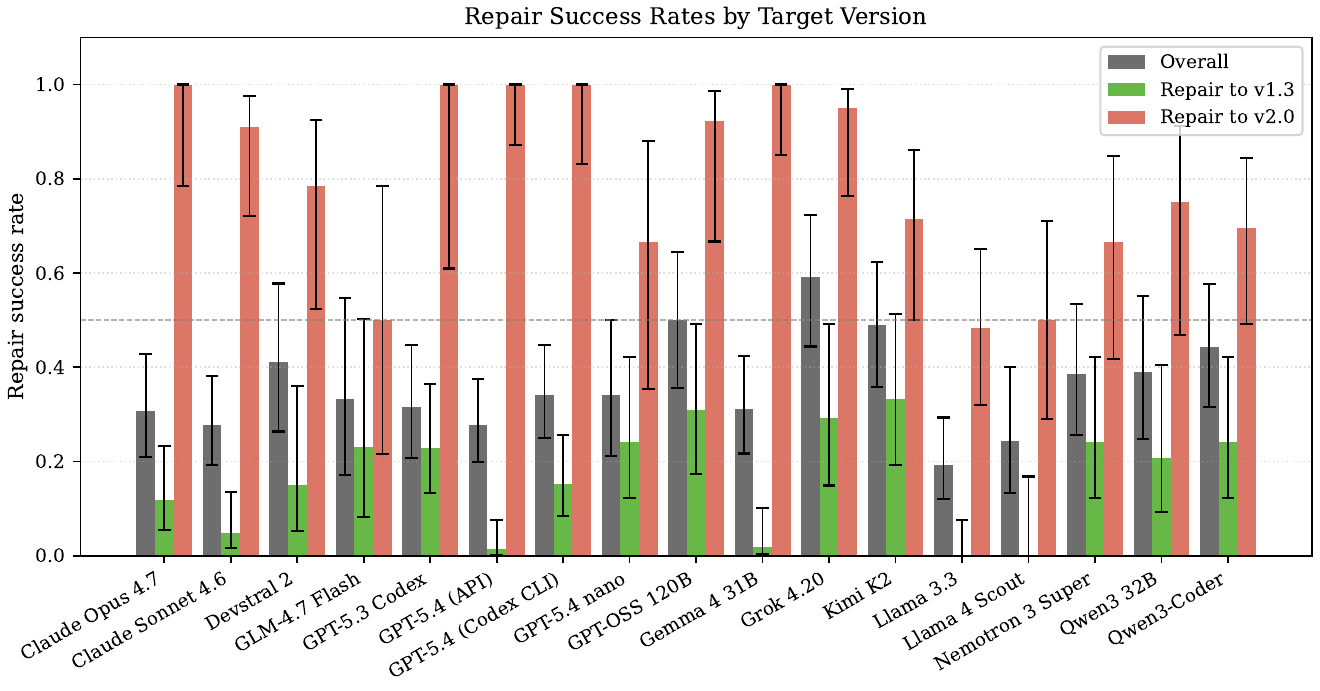}
  \caption{Repair success rates for all models: overall (gray), repair targeting
    v1.3 (green), and repair targeting v2.0 (red). Models are ordered left-to-right
    by their index in the evaluation suite. Note the consistently higher success
    rates for v2.0 repairs.}
  \label{fig:repair_rates}
\end{figure}

Figure~\ref{fig:repair_rates} visualizes the repair rates in grouped bar form.
The near-zero v1.3 repair rates for several models contrast with high v2.0
rates, reinforcing the conclusion that v1.3 represents a uniquely difficult
repair target.

\subsection{Pass@1 vs.\ Pass@3}

\begin{figure}[htbp]
  \centering
  \includegraphics[width=0.55\textwidth]{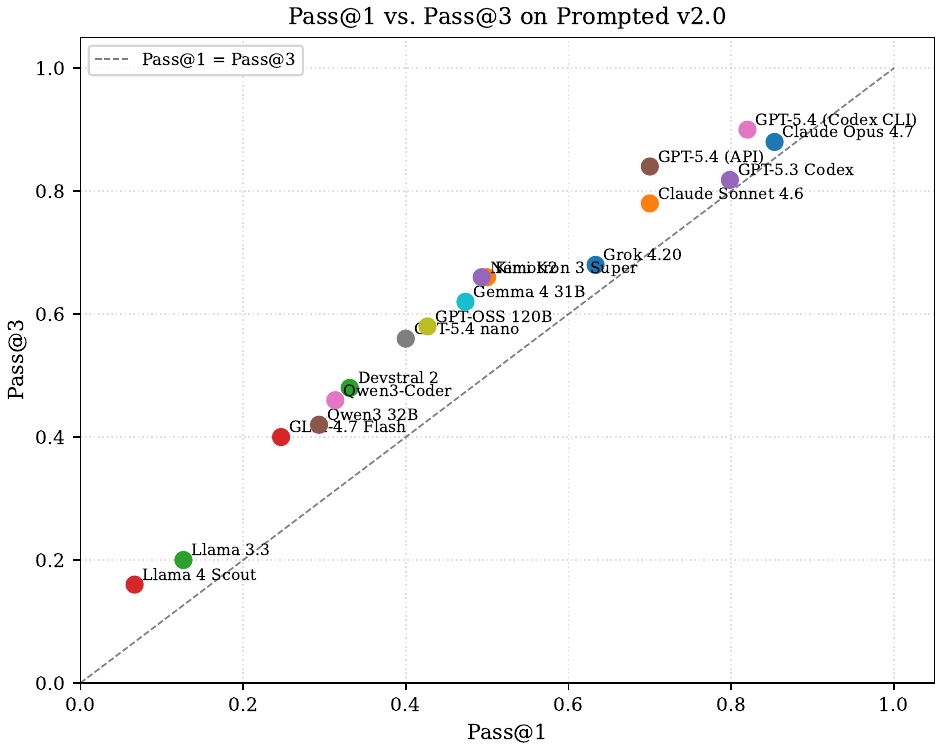}
  \caption{Pass@1 vs.\ Pass@3 for each model on the prompted v2.0, executed v2.0
    cell. Points above the dashed diagonal indicate benefit from sampling $k > 1$.
    All models gain from additional samples.}
  \label{fig:pass1_vs_pass3}
\end{figure}

Figure~\ref{fig:pass1_vs_pass3} shows the relationship between Pass@1 and
Pass@3 on the v2.0 $\to$ v2.0 cell for all models. All models lie above the
diagonal, confirming that sampling three independent candidates reliably
increases the probability of at least one success. The improvement is
especially large for GLM 4.7 Flash ($\passk{1} = 0.25$, $\passk{3} = 0.40$),
indicating that while the model rarely generates correct code on the first
attempt, its outputs exhibit diversity sufficient to include correct solutions
with higher sampling. Claude Opus 4.7 lies closest to the diagonal
($\passk{1} = 0.85$, $\passk{3} = 0.88$), reflecting both high baseline
quality and lower output variance. GPT-5.4 (Codex CLI) and GPT-5.4 (API) both
show meaningful gains ($\passk{3}$ of $0.90$ and $0.84$ respectively).

\subsection{Task Difficulty by Category}

The benchmark also allows a task-level view of where difficulty comes from. The
50 tasks span eight functional categories with clear difficulty differences
(Table~\ref{tab:task_difficulty}). Pass@1 is aggregated across models and
execution environments to give a coarse category-level view.

\begin{table}[htbp]
  \centering
  \caption{Task difficulty by category: mean Pass@1 aggregated across all
    models and execution environments. Min and max show the range across
    individual problems within each category. The final three columns show
    mean fidelity (prompt version = execution version) broken out by Qiskit
    version, revealing where the v1.3 difficulty is concentrated.}
  \label{tab:task_difficulty}
  \setlength{\tabcolsep}{4.5pt}
  \begin{tabular}{lrccc|ccc}
    \toprule
                                 &            &               &              &              & \multicolumn{3}{c}{\textbf{Fidelity by version}}                                 \\
    \textbf{Category}            & \textbf{N} & \textbf{Mean} & \textbf{Min} & \textbf{Max} & \textbf{v0.43}                                   & \textbf{v1.3} & \textbf{v2.0} \\
    \midrule
    Serialization (QASM)         & 4          & .571          & .492         & .678         & .554                                             & .554          & .652          \\
    DAG \& Circuit Manip.        & 5          & .523          & .265         & .784         & .682                                             & .380          & .553          \\
    Circuit Construction         & 6          & .492          & .266         & .743         & .634                                             & .301          & .608          \\
    Synthesis \& Controlled Ops. & 7          & .480          & .209         & .699         & .546                                             & .235          & .675          \\
    Operators \& Quantum Info    & 7          & .377          & .052         & .756         & .406                                             & .353          & .381          \\
    Transpilation \& Passes      & 8          & .347          & .091         & .682         & .393                                             & .203          & .452          \\
    Quantum Algorithms           & 8          & .309          & .018         & .819         & .382                                             & .197          & .348          \\
    Execution \& Simulation      & 5          & .122          & .013         & .391         & .316                                             & .008          & .186          \\
    \bottomrule
  \end{tabular}
\end{table}

Execution \& Simulation is the hardest category overall (mean $0.122$), with
near-zero v1.3 fidelity ($0.008$) despite higher scores on v0.43 ($0.316$) and
v2.0 ($0.186$). Serialization (QASM) is the easiest and most stable category
(mean $0.571$), with a narrow matched-version range from $0.554$ to $0.652$.
Quantum Algorithms is the second-hardest category (mean $0.309$), followed by
Transpilation \& Passes (mean $0.347$); both show substantial v1.3 weakness.
These results support a simple interpretation: tasks that depend on runtime
execution APIs and higher-level algorithm workflows are much less stable than
serialization and lower-level circuit-manipulation tasks.

\section{Analysis by Model Type}
\label{sec:model_type}

The evaluated models differ in access policy, architecture, active parameter
count, training emphasis, and knowledge cutoff. We summarize these comparisons
with mean version fidelity, computed as the average of the three diagonal
Pass@1 values.

\subsection{Closed vs.\ Open-Weight Models}

We now turn from aggregate benchmark results to model-level comparisons. With
seven closed and ten open-weight models evaluated under matched REST-API
conditions, plus one Codex CLI reference, the suite is broad enough to support
several descriptive comparisons across access policy, cutoff recency, and model
family. Table~\ref{tab:type_summary} summarizes average version fidelity.

The first pattern is a clear but incomplete closed/open gap. The six closed REST
models average mean fidelity $0.539$, while the ten open-weight REST models
average $0.314$. The gap appears on every benchmark version, but the magnitude
changes by target. On v2.0 the strongest closed models cluster between $0.633$
and $0.853$, whereas the open-weight group ranges from $0.067$ to $0.500$.
On v1.3, however, the separation narrows because most models struggle: Grok~4.20
reaches $0.513$, Kimi~K2 reaches $0.420$, and Nemotron~3~Super reaches $0.356$,
which places several open-weight models near or above some closed alternatives.
At the lower end, GPT-5.4 Nano overlaps with multiple open-weight models, so
access policy alone does not explain all variance. The evidence therefore points
to a closed-model advantage in this benchmark, but not to a strict boundary
between the two groups.

\begin{table}[htbp]
  \centering
  \caption{Mean version fidelity (Pass@1, averaged over v0.43/v1.3/v2.0
    diagonal cells) per model, with architecture and active parameter count.
    Values in parentheses show the range across individual versions.
    $^\dagger$Hybrid LatentMoE with interleaved Mamba-2 and attention layers.
    $^\ddagger$Codex CLI reference (not directly comparable to REST-API models).}
  \label{tab:type_summary}
  \setlength{\tabcolsep}{4pt}
  \begin{tabular}{llrlc}
    \toprule
    \textbf{Model}                 & \textbf{Arch.} & \textbf{Active Params} & \textbf{Focus} & \textbf{Mean Fidelity (range)} \\
    \midrule
    GPT-5.4 (Codex CLI)$^\ddagger$ & -              & undisclosed            & General        & .582 \quad (.253-.820)         \\
    \midrule
    Claude Opus 4.7                & -              & undisclosed            & General        & .669 \quad (.433-.853)         \\
    GPT-5.3-codex                  & -              & undisclosed            & Coding         & .601 \quad (.358-.798)         \\
    Grok 4.20                      & -              & undisclosed            & General        & .578 \quad (.513-.633)         \\
    GPT-5.4 (API)                  & -              & undisclosed            & General        & .553 \quad (.267-.700)         \\
    Claude Sonnet 4.6              & -              & undisclosed            & General        & .507 \quad (.213-.700)         \\
    Kimi K2                        & MoE            & 32B                    & Coding         & .496 \quad (.420-.567)         \\
    Nemotron 3 Super               & MoE$^\dagger$  & 12B                    & General        & .445 \quad (.356-.493)         \\
    GPT-OSS 120B                   & Dense          & 120B                   & General        & .430 \quad (.347-.517)         \\
    Gemma 4 31B                    & Dense          & 31B                    & General        & .344 \quad (.133-.473)         \\
    Qwen3 Coder                    & Dense          & undisclosed            & Coding         & .338 \quad (.287-.413)         \\
    Devstral 2                     & Dense          & 123B                   & Coding         & .329 \quad (.288-.369)         \\
    GPT-5.4 Nano                   & -              & undisclosed            & General        & .329 \quad (.227-.400)         \\
    Qwen3 32B                      & Dense          & 32B                    & General        & .284 \quad (.187-.373)         \\
    GLM 4.7 Flash                  & MoE            & 3B                     & Coding         & .229 \quad (.207-.247)         \\
    Llama 3.3                      & Dense          & 70B                    & General        & .173 \quad (.040-.353)         \\
    Llama 4 Scout                  & MoE            & 17B                    & General        & .073 \quad (.020-.133)         \\
    \bottomrule
  \end{tabular}
\end{table}

\subsection{Training Cutoff and Temporal Alignment}
\label{sec:cutoff}

Qiskit v1.3 was released on November 28, 2024, and Qiskit v2.0 on March 31,
2025. A model whose training data ends before these dates cannot have direct
knowledge of the corresponding API surfaces. Table~\ref{tab:cutoffs} shows the
three resulting temporal tiers for models in our expanded suite.

\begin{table}[htbp]
  \centering
  \caption{Temporal alignment between model training cutoffs and Qiskit release
    dates. Models in Tier~A have full exposure to all three benchmark versions;
    Tier~B models missed v2.0; Tier~C models missed both v1.3 and v2.0.}
  \label{tab:cutoffs}
  \setlength{\tabcolsep}{5pt}
  \begin{tabular}{llccc}
    \toprule
    \textbf{Model}      & \textbf{K.C.} & \textbf{v0.43?} & \textbf{v1.3?} & \textbf{v2.0?}              \\
    \midrule
    \multicolumn{5}{l}{\textit{Tier A - cutoff after v2.0 (Mar 2025): full SDK coverage}}                \\
    GPT-5.4 (Codex CLI) & Aug 2025      & \checkmark      & \checkmark     & \checkmark                  \\
    Claude Opus 4.7     & Apr 2025      & \checkmark      & \checkmark     & \checkmark                  \\
    Claude Sonnet 4.6   & Apr 2025      & \checkmark      & \checkmark     & \checkmark                  \\
    GPT-5.3-codex       & Aug 2025      & \checkmark      & \checkmark     & \checkmark                  \\
    GPT-5.4 (API)       & Aug 2025      & \checkmark      & \checkmark     & \checkmark                  \\
    GPT-5.4 Nano        & Aug 2025      & \checkmark      & \checkmark     & \checkmark                  \\
    Grok 4.20           & Apr 2025      & \checkmark      & \checkmark     & \checkmark                  \\
    Kimi K2             & Apr 2025      & \checkmark      & \checkmark     & \checkmark                  \\
    Nemotron 3 Super    & Mar 2025      & \checkmark      & \checkmark     & $\sim$                      \\
    \midrule
    \multicolumn{5}{l}{\textit{Tier B - cutoff after v1.3 (Nov 2024) but before v2.0: partial coverage}} \\
    Gemma 4 31B         & Jan 2025      & \checkmark      & \checkmark     & $\times$                    \\
    GPT-OSS 120B        & ?             & \checkmark      & \checkmark?    & ?                           \\
    Devstral 2          & ?             & \checkmark      & \checkmark?    & ?                           \\
    GLM 4.7 Flash       & ?             & \checkmark      & \checkmark?    & ?                           \\
    Qwen3 Coder         & ?             & \checkmark      & \checkmark?    & $\times$?                   \\
    \midrule
    \multicolumn{5}{l}{\textit{Tier C - cutoff before v1.3: no direct exposure to v1.3 or v2.0}}         \\
    Qwen3 32B           & Oct 2024      & \checkmark      & $\times$       & $\times$                    \\
    Llama 4 Scout       & Aug 2024      & \checkmark      & $\times$       & $\times$                    \\
    Llama 3.3           & Dec 2023      & \checkmark      & $\times$       & $\times$                    \\
    \bottomrule
    \multicolumn{5}{l}{\small $\sim$~borderline (released same month as v2.0); ? = estimated or undisclosed.}
  \end{tabular}
\end{table}

Temporal alignment remains informative, especially for v2.0. Tier~A models,
which plausibly include the v2.0 release in training data, generally score
higher on v2.0 than Tier~C models, which predate both v1.3 and v2.0. The
pattern is not deterministic, since models within Tier~A still range from 0.40
to 0.85 on v2.0, but the tiering is directionally consistent with the results.
In contrast, the v1.3 column shows that recency alone is not enough: even among
Tier~A models, scores range from $0.213$ to $0.513$. This weakens any claim that
training cutoff by itself explains the middle-version difficulty. A more careful
reading is that cutoff helps most when the target release corresponds to a clear
and recent API endpoint, while intermediate releases still require the model to
select the correct local idiom among overlapping patterns. Grok~4.20 is the
only Tier~A model that clears $0.50$ on v1.3, which makes it an outlier within
the high-cutoff group.

GPT-5.4 (Codex CLI) is treated as a reference rather than a direct comparator
due to its Codex CLI interface and unrestricted output tokens (see
Section~\ref{sec:methodology}). Its Tier~A placement therefore illustrates what
is achievable with full temporal coverage under a different interface, but it
should not be used as evidence for a like-for-like API comparison.

\subsection{Dense vs.\ MoE Architecture}

Dense and MoE open-weight models are interleaved throughout the ranking, so the
results do not support a clean architectural advantage for either family.
Kimi~K2, a MoE model with 32B active parameters, leads the open-weight group,
while Llama~4~Scout, also MoE, is last. Dense models likewise span a wide range,
from GPT-OSS~120B at $0.430$ mean fidelity to Llama~3.3 at $0.173$. This spread
within each architectural family is larger than the apparent gap between the two
families. The more consistent signal is therefore overall model quality and, to
a lesser extent, effective active capacity, rather than Dense versus MoE alone.

\subsection{Training Focus: General vs.\ Code-Oriented}

Coding focus alone is not enough. Kimi~K2 performs well, but other code-oriented
models such as Devstral~2, Qwen3~Coder, and GLM~4.7 Flash are much weaker.
Qwen3~Coder only modestly improves on Qwen3~32B, which suggests that general
coding specialization does not automatically transfer to quantum SDK knowledge.
At the same time, GPT-5.3-codex outperforms several general-purpose models,
which indicates that coding-focused adaptation can help when it is paired with
stronger underlying model capability and sufficiently recent knowledge. The
results therefore support a qualified claim: code orientation is beneficial for
some models, but it is not a reliable proxy for version-robust quantum SDK use.

\subsection{The Scale vs.\ Specialization Trade-off}

Scale matters, but only loosely. Nemotron~3~Super performs better than several
larger dense models despite a smaller active parameter count, whereas Llama~4
Scout has high total parameters but the lowest open-weight fidelity. Within the
GPT-5.4 family, the Nano variant also trails the larger model, which suggests
that scale still contributes within a shared training period.

\subsection{Repair Effectiveness by Model Type}

Repair effectiveness broadly tracks baseline competence, but only up to a point.
Stronger models usually repair more successfully overall, yet the target version
matters even more than the model family. For example, Claude Opus 4.7 and
GPT-5.4 repair to v2.0 perfectly in the attempted cases, but both are weak on
repair to v1.3. Several open-weight models show the same asymmetry. This means
repair is not simply a continuation of baseline generation quality. Instead, the
evidence suggests that documentation helps most when it can map an outdated
pattern onto a cleaner successor API, and helps less when the target version
requires choosing among partially overlapping interfaces.

\section{Discussion}
\label{sec:discussion}

\subsection{The v1.3 Anomaly}

The results point to one dominant theme: v1.3 is unusually difficult. This
reading is supported independently by the fidelity results, the drift heatmaps,
and the repair table. Among the 16 matched
REST models, every model is below $0.52$ on the v1.3 diagonal, and several are
below $0.25$. Repair toward v1.3 is similarly poor for many otherwise strong
models. Why v1.3 is harder cannot be established from our benchmark alone, but
the evidence is consistent with three non-exclusive possibilities: the version
sits between older and newer idioms, some models had limited direct exposure to
it, and a simple prompt anchor may not be enough to force the correct
intermediate API style. The key point is that all three explanations are
compatible with the reported data, whereas a stronger causal claim would not be
justified. Grok~4.20 is the main exception and deserves follow-up on a larger
benchmark because it is the only REST model above $0.50$ on v1.3.

\subsection{Two Failure Regimes}

The error analysis suggests two practical failure regimes. Some models fail
early with broken imports and symbol references, as seen in the high
\texttt{ImportError} counts for Llama~4~Scout, Llama~3.3, Qwen3~32B, and other
lower-fidelity systems. Others generate code that nearly works but uses
deprecated interfaces, which appears as a larger share of
\texttt{DeprecationWarning} in stronger models such as Claude~Opus~4.7,
GPT-5.4, and Claude~Sonnet~4.6. This distinction matters because the second
regime is more repairable: it preserves more of the required program structure,
so migration notes can often guide the model to a working update. The evidence
here supports a difference in failure stage, not a claim about internal model
reasoning.

\subsection{Repair Asymmetry: v1.3 vs.\ v2.0}

The repair asymmetry is also clear. Migration-note prompting works much better
for v2.0 than for v1.3, and this pattern holds across both strong closed models
and several open-weight models. One plausible explanation is that migration to v2.0 is often easier to state
explicitly because it completes removals that had already been signaled in the
v1.x line, whereas repair toward v1.3 may require choosing among partially
transitioned interfaces that can all appear locally plausible. That mechanism is
not measured directly here, but it is consistent with the repair table and with
the error-profile shift toward deprecation-heavy failures in stronger models.

\subsection{Task Category and Hardness}

Execution-heavy tasks account for much of the benchmark difficulty. The category
summary shows that Execution \& Simulation is the hardest group overall and is
especially weak on v1.3. Serialization remains comparatively stable, and
circuit construction plus DAG/circuit manipulation are substantially stronger
overall than execution-heavy and algorithmic workflows, although they still show
v1.3 dips. This matters in practice because not all quantum programming
workflows depend equally on the most volatile parts of the SDK. A user who asks
for serialization or lower-level circuit manipulation may see moderate
robustness across versions, whereas a user who asks for backend execution,
sampler-style workflows, or higher-level algorithms is much more likely to hit
API drift.

\subsection{Practical Implications}

The practical message is straightforward. The best models in our study are often
useful, but even strong performance on aggregate quantum-code generation does
not guarantee version alignment. Users who need executable code should therefore
treat the SDK version as part of the task specification, not as incidental
context, and should verify generated outputs in the intended environment. This
is especially important for execution and simulation workflows, which are the
least stable categories in our benchmark. The repair results suggest a second
lesson: documentation-guided patching can be effective, but its benefit depends
strongly on the migration target. In our setting, migration guidance is much
more reliable for repair to v2.0 than for repair to v1.3, so intermediate
version targets may still require substantial manual debugging. More broadly,
these findings suggest that future evaluations of quantum code generation should
report SDK version explicitly. Without that information, some failures that look
like reasoning errors may in fact be version mismatches.

The comparison between GPT-5.4 (API) and GPT-5.4 (Codex CLI) also suggests
that benchmark outcomes may depend on deployment pathway, not only on the
nominal model family. Although both are GPT-5.4-based systems, the Codex
CLI setting exhibits a markedly different version profile, with strong
performance on v0.43 and v2.0, weak alignment to v1.3, and much stronger
repair performance when targeting v2.0 than v1.3. We do not treat this as
evidence of a purely model-intrinsic difference, since interface conditions
and tooling context may also contribute. However, the result reinforces a
broader lesson of this study: version fidelity should be evaluated
empirically for the exact deployment setting in which a coding system will
be used.

\section{Limitations}
\label{sec:limitations}

Several limitations affect how these results should be interpreted.

\textbf{API-level validity only.} Our test harnesses verify that generated code
executes without exception and returns a non-null result, but do not verify
semantic circuit correctness (e.g., correct gate sequences, expected measurement
distributions). A circuit that passes our tests may still implement the wrong
quantum operation.

\textbf{SDK-agnostic rewriting trade-offs.} The 50 tasks are adapted from the
Qiskit HumanEval benchmark~\cite{vishwakarma2024qiskithumaneval} with test harnesses rewritten
to avoid version-specific API calls. This ensures that test execution does not
itself depend on the SDK version under test, but it means the benchmark is not
identical to the original Qiskit HumanEval and semantic correctness for some
tasks cannot be guaranteed. Results should be interpreted as measuring
API-level generation quality rather than circuit correctness.

\textbf{Training cutoff confound.} We cannot cleanly separate model capability
from training-data recency. A stronger result on v2.0 may reflect either better
reasoning or simply later exposure to the API.

\textbf{GPT-5.4 comparability.} GPT-5.4 (Codex CLI) was evaluated under a
different deployment pathway from the matched REST-API setup, including
differences in interface conditions and output-budget policy. Therefore, its
results are not directly comparable to the REST-API results as a controlled
apples-to-apples model comparison. We retain it as an informative reference
point, but any observed differences should be interpreted as deployment-conditioned
effects rather than purely model-intrinsic ones.

\textbf{Migration note quality.} Repair effectiveness is partly a function of
the quality of the hand-curated migration notes. Higher-quality, more targeted
migration summaries could improve repair rates beyond what we observe.

\textbf{Single temperature.} All experiments use $T = 0.8$. A temperature
sweep could reveal whether lower temperatures improve fidelity at the cost of
output diversity.

\textbf{No iterative repair.} We attempt repair once per failure. Iterative
repair (executing, observing the new error, re-attempting) could substantially
improve success rates, particularly for the v1.3 target where a single pass
rarely suffices.

\textbf{Conda isolation.} We use Conda environments for version isolation.
Docker containers would provide stricter isolation, particularly for system
library dependencies.

\section{Conclusion}
\label{sec:conclusion}

We introduced \textsc{quantum-api-drift}, a benchmark for measuring API drift in
LLM-generated quantum SDK code, and instantiated it with Qiskit across v0.43,
v1.3, and v2.0. Across 17 evaluated models, the central result is that version
alignment is a distinct challenge from general code generation competence.
Models that can often produce plausible quantum programs still fail regularly
when asked to target a specific SDK release.

Three conclusions stand out. First, API drift is large and structured: among the
16 matched REST models, version fidelity ranges from 0.02 to 0.85, and the full
drift matrices show substantial asymmetry across version transitions. Second,
the benchmark versions are not equally difficult. Qiskit v1.3 is the weakest
target in both direct generation and repair, which suggests that intermediate
API epochs deserve special attention in future evaluation. Third, failure modes
and repairability are informative in their own right. We observe a broad shift
from broken-import failures in weaker models to deprecation-heavy failures in
stronger models, and we find that documentation-guided repair is only partly
effective overall and much more successful for migration to v2.0 than to v1.3.

These findings motivate a broader evaluation agenda for quantum code generation.
Future work should combine version-aware benchmarking with stronger semantic
correctness checks, iterative repair, retrieval from live SDK documentation, and
controlled studies that isolate the effect of interface, tooling, and output
budget within the same model family. An important next step is to extend the
framework beyond Qiskit to additional quantum SDKs such as Cirq, PennyLane, and
CUDA-Q in order to test whether the same drift patterns generalize.

To support reproducibility and follow-up work, we release the benchmark
framework, task set, execution harnesses, and experimental results at
\url{https://github.com/arasyi/quantum-api-drift}.

\bibliographystyle{plain}
\bibliography{references}

\appendix
\clearpage
\section{Appendix: Complete Experimental Results}

This appendix provides the complete visualization set and compact summary tables for the results discussed in the main text. Machine-readable CSV versions of the full drift matrix, compatibility summaries, repair summaries, and error taxonomies are generated alongside these figures.

\subsection{All-model drift matrices}
\begin{figure}[h]
    \centering
    \includegraphics[width=\textwidth]{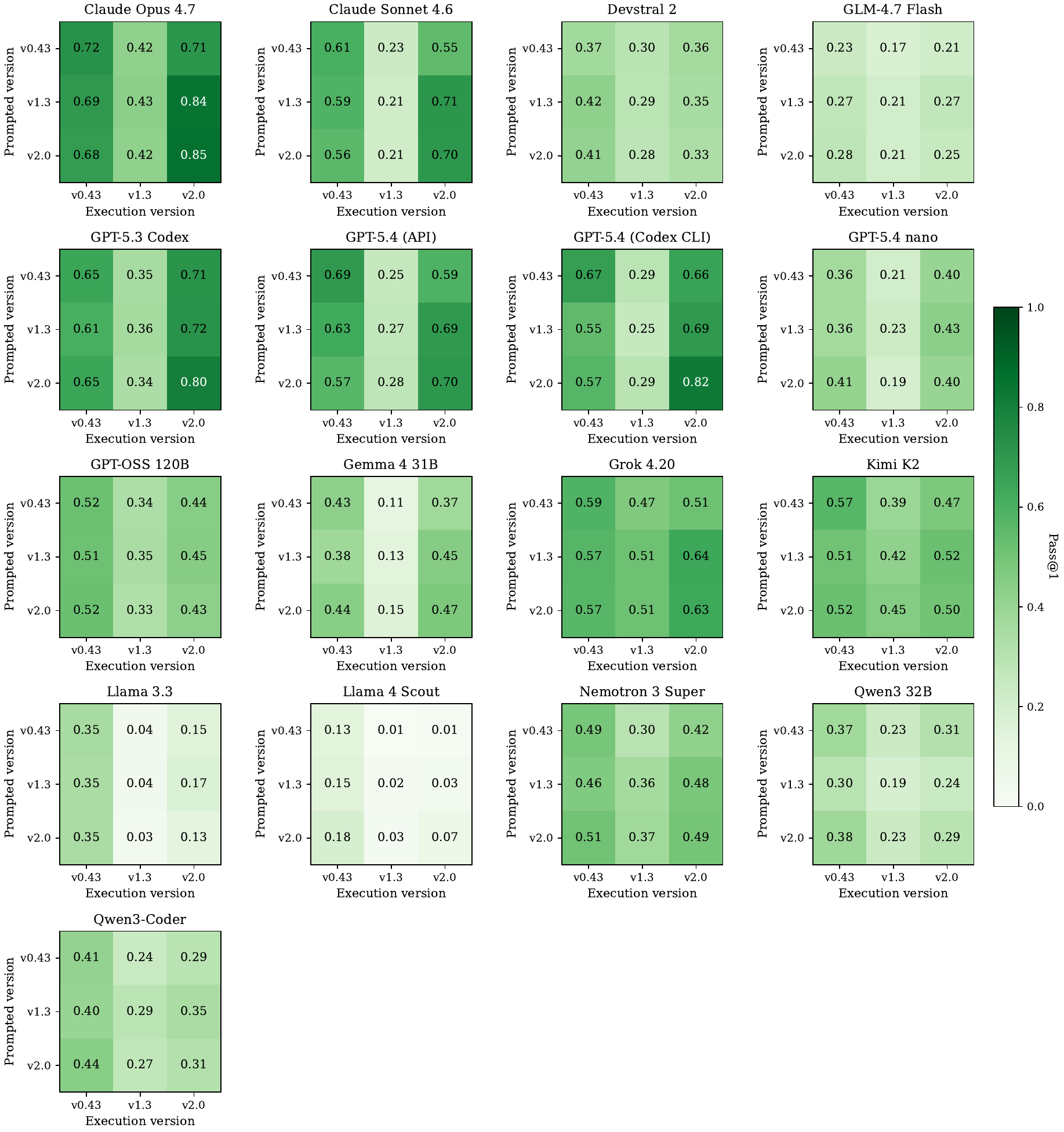}
    \caption{Complete drift heatmaps for all evaluated models Darker cells indicate higher Pass@1. Diagonal cells correspond to version fidelity, while off-diagonal cells capture cross-version transfer.}
    \label{fig:appendix-heatmaps-all}
\end{figure}

Figures~\ref{fig:appendix-heatmaps-all} show the full $3 \times 3$ drift matrix for every model. Each heatmap reports Pass@1 for the prompted version and execution version pairing, making it possible to inspect diagonal fidelity, forward transfer, and backward transfer for all evaluated systems.

\clearpage
\subsection{Version fidelity and compatibility summaries}
Table~\ref{tab:appendix-version-fidelity} summarizes diagonal performance for each model on Qiskit v0.43, v1.3, and v2.0. Table~\ref{tab:appendix-compatibility} aggregates the same results into mean diagonal fidelity, mean forward off-diagonal performance, mean backward off-diagonal performance, and the largest observed forward drop.

\begin{table}[h]
    \centering
    \caption{Version fidelity summary across prompted and execution versions. Mean scores are averages of the diagonal cells in the model\textquotesingle s drift matrix.}
    \label{tab:appendix-version-fidelity}
    \scriptsize
    \setlength{\tabcolsep}{5pt}
    \begin{tabular}{lcccccccc}
        \toprule
        Model               & v0.43 P@1 & v1.3 P@1 & v2.0 P@1 & Mean P@1 & v0.43 P@3 & v1.3 P@3 & v2.0 P@3 & Mean P@3 \\
        \midrule
        Claude Opus 4.7     & 0.72      & 0.4333   & 0.8533   & 0.6689   & 0.76      & 0.46     & 0.88     & 0.7      \\
        GPT-5.3 Codex       & 0.6471    & 0.3577   & 0.7984   & 0.601    & 0.7083    & 0.4255   & 0.8182   & 0.6507   \\
        GPT-5.4 (Codex CLI) & 0.6733    & 0.2533   & 0.82     & 0.5822   & 0.76      & 0.34     & 0.9      & 0.6667   \\
        Grok 4.20           & 0.5867    & 0.5133   & 0.6333   & 0.5778   & 0.64      & 0.56     & 0.68     & 0.6267   \\
        GPT-5.4 (API)       & 0.6933    & 0.2667   & 0.7      & 0.5533   & 0.8       & 0.32     & 0.84     & 0.6533   \\
        Claude Sonnet 4.6   & 0.6067    & 0.2133   & 0.7      & 0.5067   & 0.68      & 0.24     & 0.78     & 0.5667   \\
        Kimi K2             & 0.5667    & 0.42     & 0.5      & 0.4956   & 0.76      & 0.54     & 0.66     & 0.6533   \\
        Nemotron 3 Super    & 0.4867    & 0.3557   & 0.4933   & 0.4452   & 0.64      & 0.52     & 0.66     & 0.6067   \\
        GPT-OSS 120B        & 0.5168    & 0.3467   & 0.4267   & 0.43     & 0.66      & 0.46     & 0.58     & 0.5667   \\
        Gemma 4 31B         & 0.4267    & 0.1333   & 0.4733   & 0.3444   & 0.52      & 0.16     & 0.62     & 0.4333   \\
        Qwen3-Coder         & 0.4133    & 0.2867   & 0.3133   & 0.3378   & 0.58      & 0.38     & 0.46     & 0.4733   \\
        Devstral 2          & 0.3691    & 0.2877   & 0.3311   & 0.3293   & 0.58      & 0.42     & 0.48     & 0.4933   \\
        GPT-5.4 nano        & 0.36      & 0.2267   & 0.4      & 0.3289   & 0.52      & 0.36     & 0.56     & 0.48     \\
        Qwen3 32B           & 0.3733    & 0.1867   & 0.2933   & 0.2844   & 0.5       & 0.3      & 0.42     & 0.4067   \\
        GLM-4.7 Flash       & 0.2333    & 0.2067   & 0.2467   & 0.2289   & 0.52      & 0.36     & 0.4      & 0.4267   \\
        Llama 3.3           & 0.3533    & 0.04     & 0.1267   & 0.1733   & 0.46      & 0.06     & 0.2      & 0.24     \\
        Llama 4 Scout       & 0.1333    & 0.02     & 0.0667   & 0.0733   & 0.22      & 0.04     & 0.16     & 0.14     \\
        \bottomrule
    \end{tabular}
\end{table}

\begin{table}[h]
    \centering
    \caption{Compatibility summary across models. Mean forward and backward scores average off-diagonal cells in the corresponding direction. Largest forward drop is the biggest decrease from a diagonal cell to a later-version execution target.}
    \label{tab:appendix-compatibility}
    \scriptsize
    \setlength{\tabcolsep}{5pt}
    \begin{tabular}{lccccr}
        \toprule
        Model               & Mean diagonal & Mean forward & Mean backward & Largest forward pair & Drop   \\
        \midrule
        Claude Opus 4.7     & 0.6689        & 0.6556       & 0.5978        & v0.43->v1.3          & 0.3    \\
        GPT-5.3 Codex       & 0.601         & 0.5938       & 0.5299        & v0.43->v1.3          & 0.2941 \\
        GPT-5.4 (Codex CLI) & 0.5822        & 0.5467       & 0.4711        & v0.43->v1.3          & 0.3867 \\
        Grok 4.20           & 0.5778        & 0.54         & 0.5511        & v0.43->v1.3          & 0.12   \\
        GPT-5.4 (API)       & 0.5533        & 0.5133       & 0.4911        & v0.43->v1.3          & 0.44   \\
        Claude Sonnet 4.6   & 0.5067        & 0.4956       & 0.4533        & v0.43->v1.3          & 0.3733 \\
        Kimi K2             & 0.4956        & 0.4622       & 0.4933        & v0.43->v1.3          & 0.1733 \\
        Nemotron 3 Super    & 0.4452        & 0.4011       & 0.4477        & v0.43->v1.3          & 0.1867 \\
        GPT-OSS 120B        & 0.43          & 0.4084       & 0.4533        & v0.43->v1.3          & 0.1812 \\
        Gemma 4 31B         & 0.3444        & 0.3089       & 0.3244        & v0.43->v1.3          & 0.32   \\
        Qwen3-Coder         & 0.3378        & 0.2933       & 0.3711        & v0.43->v1.3          & 0.1733 \\
        Devstral 2          & 0.3293        & 0.3357       & 0.3735        & v0.43->v1.3          & 0.0738 \\
        GPT-5.4 nano        & 0.3289        & 0.3467       & 0.32          & v0.43->v1.3          & 0.1533 \\
        Qwen3 32B           & 0.2844        & 0.2622       & 0.3044        & v0.43->v1.3          & 0.14   \\
        GLM-4.7 Flash       & 0.2289        & 0.2178       & 0.2533        & v0.43->v1.3          & 0.06   \\
        Llama 3.3           & 0.1733        & 0.1178       & 0.2422        & v0.43->v1.3          & 0.3133 \\
        Llama 4 Scout       & 0.0733        & 0.0178       & 0.1178        & v0.43->v1.3          & 0.1267 \\
        \bottomrule
    \end{tabular}
\end{table}

\clearpage
\subsection{Sampling and repair analyses}
\begin{figure}[h]
    \centering
    \includegraphics[width=0.92\textwidth]{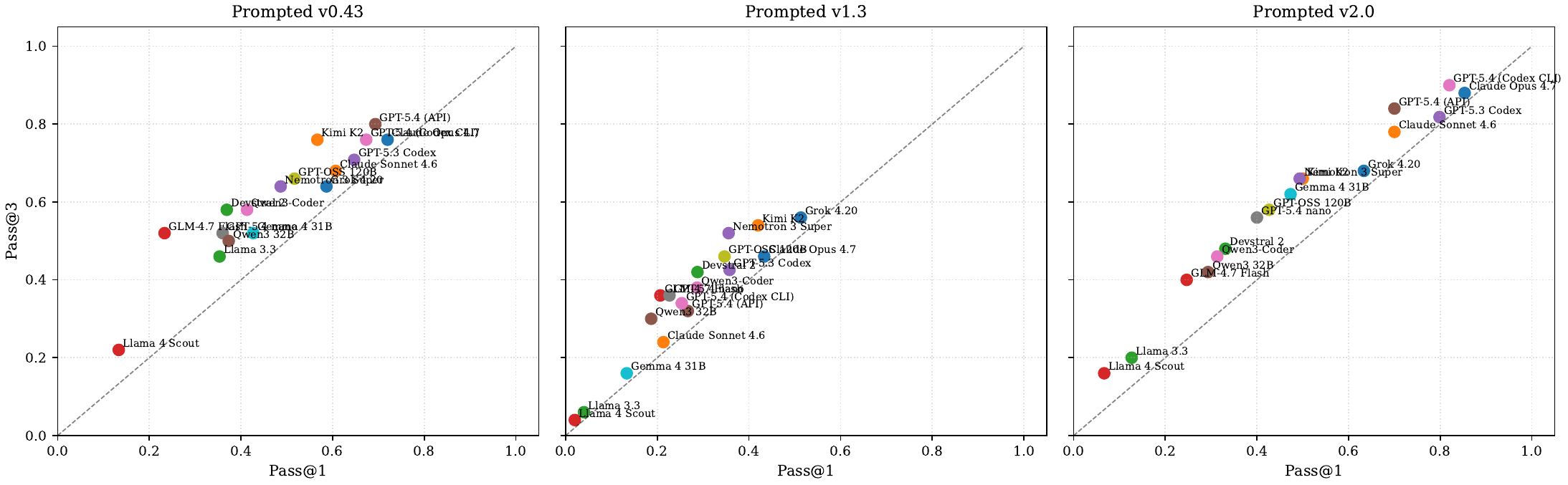}
    \caption{Pass@1 versus Pass@3 for diagonal cells across all three target versions. Points above the identity line benefit from multiple samples.}
    \label{fig:appendix-pass1-pass3}
\end{figure}

Figure~\ref{fig:appendix-pass1-pass3} compares Pass@1 and Pass@3 on the diagonal cells for all three target versions. Table~\ref{tab:appendix-repair-summary} reports overall repair success and the breakdown by v1.3 and v2.0 repair targets, including confidence intervals.

\begin{table}[h]
    \centering
    \caption{Repair summary by model. Rates include Wilson 95\% confidence intervals in brackets.}
    \label{tab:appendix-repair-summary}
    \scriptsize
    \setlength{\tabcolsep}{3pt}
    \begin{tabular}{lrrccrrccrrcc}
        \toprule
        \textbf{Model}      & \multicolumn{4}{c}{\textbf{Overall}} & \multicolumn{4}{c}{\textbf{v1}} & \multicolumn{4}{c}{\textbf{v2}}                                                                                                         \\
                            & succ.                                & total                           & rate                            & 95\% CI         & succ. & total & rate   & 95\% CI         & succ. & total & rate   & 95\% CI         \\
        \midrule
        Grok 4.20           & 26                                   & 44                              & 0.5909                          & [0.4441,0.7231] & 7     & 24    & 0.2917 & [0.1491,0.4917] & 19    & 20    & 0.95   & [0.7639,0.9911] \\
        GPT-OSS 120B        & 21                                   & 42                              & 0.5                             & [0.3553,0.6447] & 9     & 29    & 0.3103 & [0.1728,0.4923] & 12    & 13    & 0.9231 & [0.6669,0.9863] \\
        Kimi K2             & 25                                   & 51                              & 0.4902                          & [0.3586,0.6232] & 10    & 30    & 0.3333 & [0.1923,0.5122] & 15    & 21    & 0.7143 & [0.5004,0.8619] \\
        Qwen3-Coder         & 23                                   & 52                              & 0.4423                          & [0.3159,0.5766] & 7     & 29    & 0.2414 & [0.1222,0.4211] & 16    & 23    & 0.6957 & [0.4913,0.844]  \\
        Devstral 2          & 14                                   & 34                              & 0.4118                          & [0.2637,0.5778] & 3     & 20    & 0.15   & [0.0524,0.3604] & 11    & 14    & 0.7857 & [0.5241,0.9243] \\
        Qwen3 32B           & 14                                   & 36                              & 0.3889                          & [0.2478,0.5514] & 5     & 24    & 0.2083 & [0.0924,0.4047] & 9     & 12    & 0.75   & [0.4677,0.9111] \\
        Nemotron 3 Super    & 17                                   & 44                              & 0.3864                          & [0.2572,0.5338] & 7     & 29    & 0.2414 & [0.1222,0.4211] & 10    & 15    & 0.6667 & [0.4171,0.8482] \\
        GPT-5.4 nano        & 13                                   & 38                              & 0.3421                          & [0.2121,0.5011] & 7     & 29    & 0.2414 & [0.1222,0.4211] & 6     & 9     & 0.6667 & [0.3542,0.8794] \\
        GPT-5.4 (Codex CLI) & 29                                   & 85                              & 0.3412                          & [0.2492,0.4469] & 10    & 66    & 0.1515 & [0.0844,0.2569] & 19    & 19    & 1      & [0.8318,1]      \\
        GLM-4.7 Flash       & 7                                    & 21                              & 0.3333                          & [0.1719,0.5463] & 3     & 13    & 0.2308 & [0.0818,0.5026] & 4     & 8     & 0.5    & [0.2152,0.7848] \\
        GPT-5.3 Codex       & 17                                   & 54                              & 0.3148                          & [0.2068,0.4474] & 11    & 48    & 0.2292 & [0.1331,0.3654] & 6     & 6     & 1      & [0.6097,1]      \\
        Gemma 4 31B         & 23                                   & 74                              & 0.3108                          & [0.2169,0.4234] & 1     & 52    & 0.0192 & [0.0034,0.1012] & 22    & 22    & 1      & [0.8513,1]      \\
        Claude Opus 4.7     & 20                                   & 65                              & 0.3077                          & [0.2089,0.428]  & 6     & 51    & 0.1176 & [0.055,0.2338]  & 14    & 14    & 1      & [0.7847,1]      \\
        GPT-5.4 (API)       & 27                                   & 97                              & 0.2784                          & [0.1989,0.3747] & 1     & 71    & 0.0141 & [0.0025,0.0756] & 26    & 26    & 1      & [0.8713,1]      \\
        Claude Sonnet 4.6   & 23                                   & 83                              & 0.2771                          & [0.1923,0.3816] & 3     & 61    & 0.0492 & [0.0169,0.1349] & 20    & 22    & 0.9091 & [0.7218,0.9747] \\
        Llama 4 Scout       & 9                                    & 37                              & 0.2432                          & [0.1336,0.4012] & 0     & 19    & 0      & [0,0.1682]      & 9     & 18    & 0.5    & [0.2903,0.7097] \\
        Llama 3.3           & 15                                   & 78                              & 0.1923                          & [0.1202,0.2934] & 0     & 47    & 0      & [0,0.0756]      & 15    & 31    & 0.4839 & [0.3197,0.6516] \\
        \bottomrule
    \end{tabular}
\end{table}


\clearpage
\subsection{Error taxonomy by prompted version}
\begin{figure}[h]
    \centering
    \includegraphics[width=\textwidth]{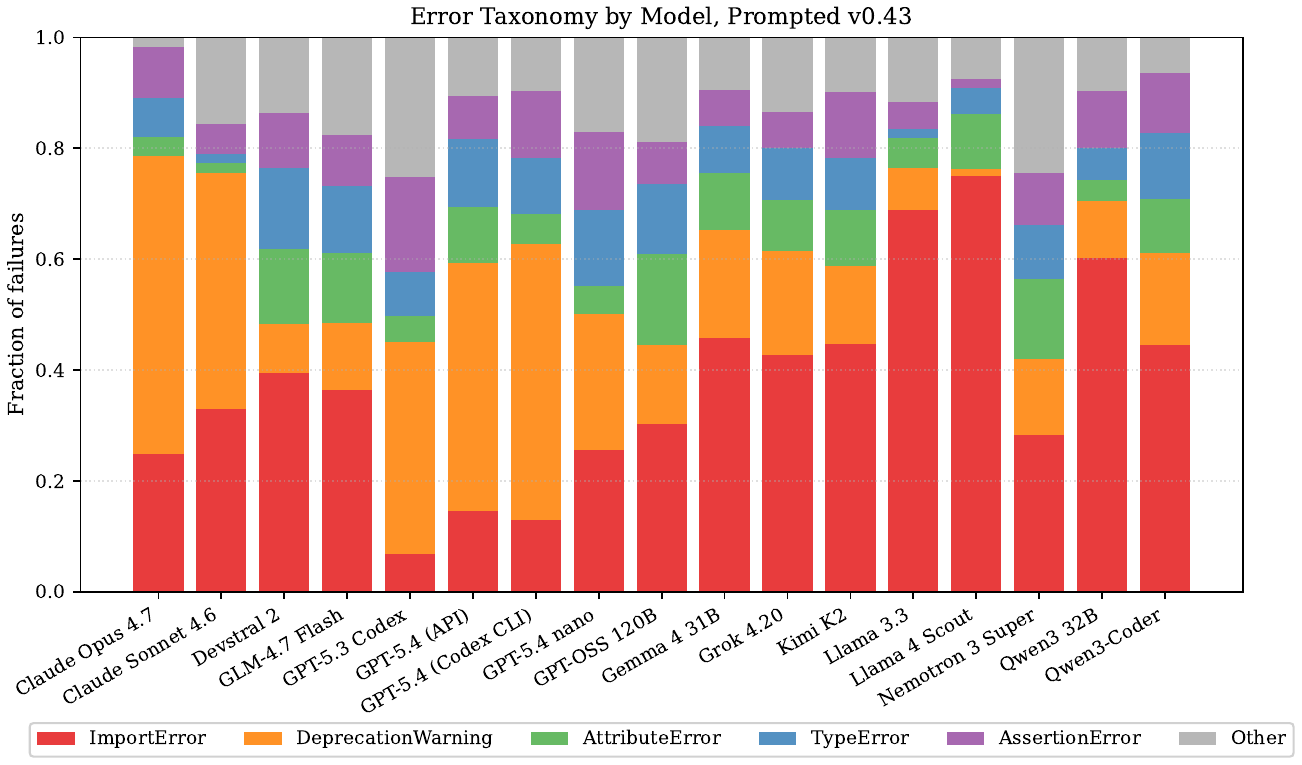}
    \caption{Normalized error taxonomy for prompts targeting Qiskit v0.43.}
    \label{fig:appendix-error-v0}
\end{figure}

\begin{figure}[h]
    \centering
    \includegraphics[width=\textwidth]{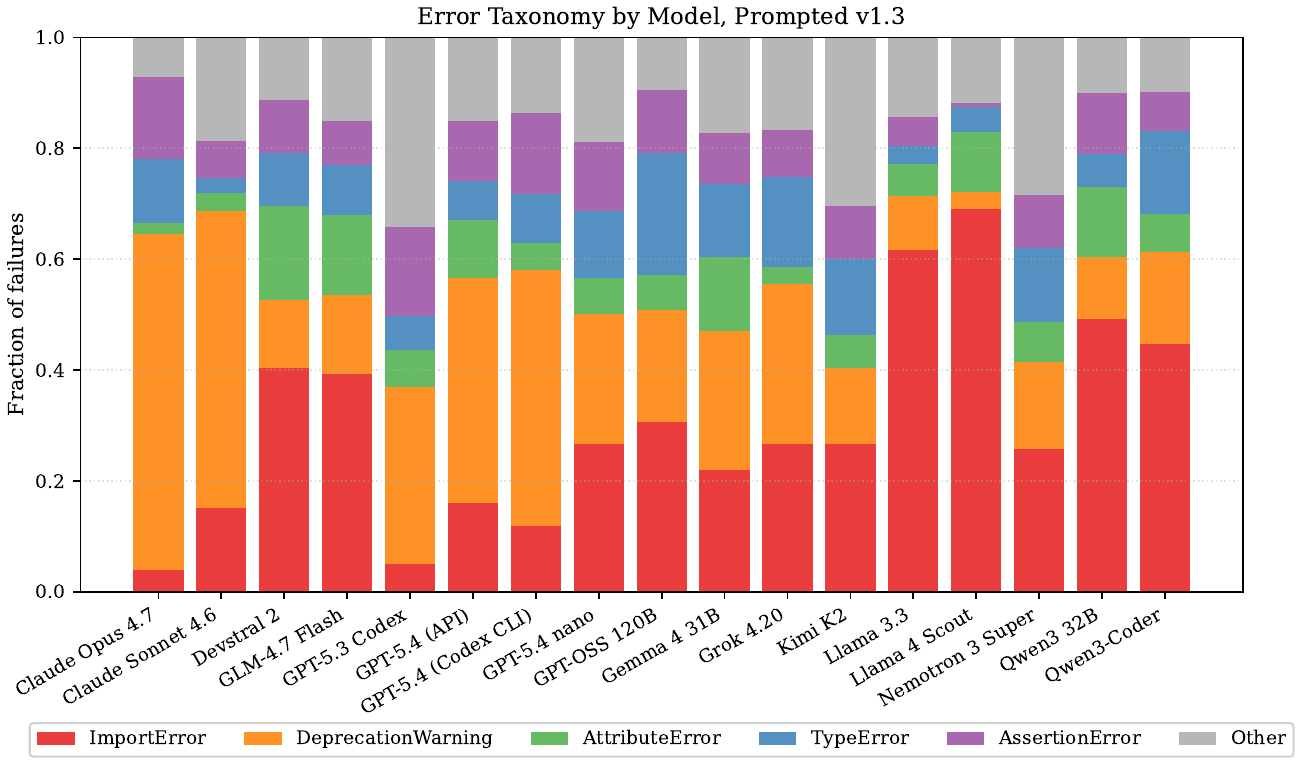}
    \caption{Normalized error taxonomy for prompts targeting Qiskit v1.3.}
    \label{fig:appendix-error-v1}
\end{figure}

\begin{figure}[h]
    \centering
    \includegraphics[width=\textwidth]{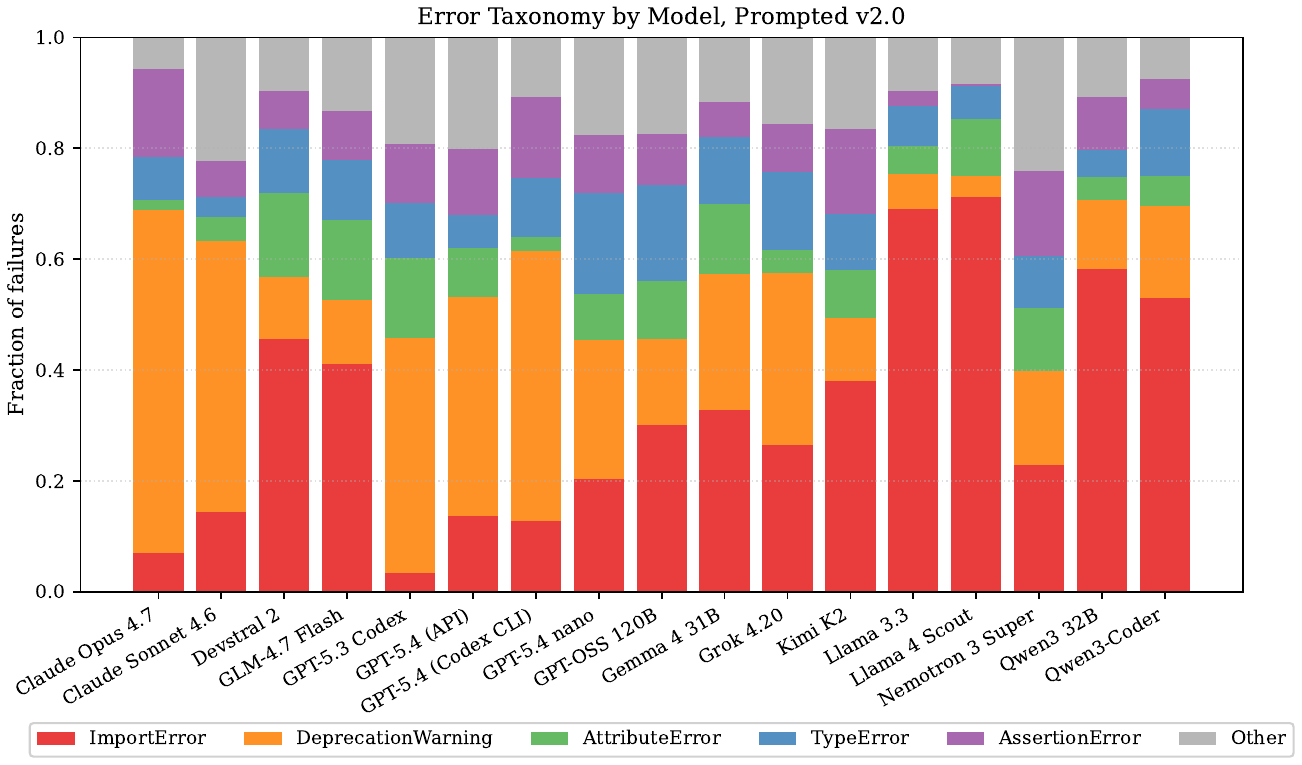}
    \caption{Normalized error taxonomy for prompts targeting Qiskit v2.0.}
    \label{fig:appendix-error-v2}
\end{figure}

Figures~\ref{fig:appendix-error-v0}--\ref{fig:appendix-error-v2} break down the normalized failure taxonomy separately for prompts targeting Qiskit v0.43, v1.3, and v2.0. This view complements the aggregate error taxonomy in the main text by revealing whether the same failure modes persist across prompted versions.

\end{document}